\documentclass[aps,prl,reprint,amsmath,amssymb,showpacs,superscriptaddress]{revtex4-2}
\usepackage[utf8]{inputenc}
\usepackage[T1]{fontenc}

\usepackage{bm}
\usepackage{graphicx}
\usepackage{float}
\usepackage{color}
\usepackage{multirow}
\usepackage{xcolor}
\usepackage{booktabs}

\usepackage[colorlinks,allcolors=blue]{hyperref}
\usepackage{orcidlink}
\makeatletter
\def\NAT@def@citea{\def\@citea{\NAT@separator}}
\makeatother

\usepackage[sectionbib]{bibunits} 

\defaultbibliographystyle{apsrev4-2} 
\defaultbibliography{ref.bib}

\begin{document}

\begin{bibunit}

\title{Universality of nucleon short-range behavior with chiral forces}

\author{Xiang-Xiang Sun\orcidlink{0000-0003-2809-4638}}
\email{x.sun@fz-juelich.de}
\affiliation{Institute for Advanced Simulation (IAS-4), Forschungszentrum J\"{u}lich, D-52425 J\"{u}lich, Germany}

\author{Hoai Le\orcidlink{0000-0003-1776-9468}}
\affiliation{Institute for Advanced Simulation (IAS-4), Forschungszentrum J\"{u}lich, D-52425 J\"{u}lich, Germany}

\author{Ulf-G. Mei{\ss}ner\orcidlink{0000-0003-1254-442X}}
\email{meissner@hiskp.uni-bonn.de}
\affiliation{Helmholtz-Institut~f\"{u}r~Strahlen-~und~Kernphysik~and~Bethe~Center~for~Theoretical~Physics, Universit\"{a}t~Bonn,~D-53115~Bonn,~Germany} 
\affiliation{Institute for Advanced Simulation (IAS-4), Forschungszentrum J\"{u}lich, D-52425 J\"{u}lich, Germany}
\affiliation{Peng Huanwu Collaborative Center for Research and Education, International Institute for Interdisciplinary and Frontiers, Beihang University, Beijing 100191, China}
\affiliation{CASA, Forschungszentrum J\"{u}lich, 52425 Ju\"{u}lich, Germany}

\author{Andreas Nogga\orcidlink{0000-0003-2156-748X}}
\thanks{Corresponding author: a.nogga@fz-juelich.de}
\affiliation{Institute for Advanced Simulation (IAS-4), Forschungszentrum J\"{u}lich, D-52425 J\"{u}lich, Germany}
\affiliation{CASA, Forschungszentrum J\"{u}lich, 52425 Ju\"{u}lich, Germany}

\date{\today}
\begin{abstract}

Modern advanced nuclear \textit{ab initio} approaches
with the similarity renormalization group (SRG) softened interactions miss high-momentum information, 
thus rendering them less suitable
for characterizing nucleon-nucleon short-range physics.
We introduce a novel framework to construct SRG-independent nuclear wave functions from No-Core Shell Model calculations. 
Applying our method to densities obtained with semilocal momentum-space-regularized chiral NN and NNN forces, we show key universalities of short-range behavior: (1) The two-body density ratio in the $np$ $S=1$ channel, relative to the deuteron ($d$), is remarkably insensitive to interaction details. (2) More strikingly, while the ratio of \textit{total} two-body densities to the deuteron exhibits cutoff dependence, the same ratio to the $\alpha$-particle (\({}^4\text{He}\)) is almost independent of the interactions.

\end{abstract}
\maketitle

{\it Introduction}---
The properties of the atomic nucleus are governed by
the interactions between nucleons constrained
by free nucleon-nucleon (NN) scattering data. These interactions can be modeled 
by, e.g., phenomenologically inspired one-boson exchange interactions \cite{Wiringa:1994wb,Machleidt:2000ge} (at least the 1$\pi$ exchange is 
usually part of this) or, in recent years, systematically using chiral effective field theory (ChEFT) \cite{Epelbaum:2008ga,Hammer:2012id,Machleidt:2011zz}.
The short-range parts of NN interactions are
connected with the repulsive core
and tensor components of nuclear forces,
which usually leads to complexities 
in modern nuclear \textit{ab initio} theories.
Isolating such short-range correlations (SRC) uniquely is impossible due to the existence of equivalent interactions that are related by unitary transformations and that alter especially the short range part. 
Their understanding within a given framework is nevertheless considered crucial for addressing fundamental questions in nuclear physics, 
including nuclear interactions \cite{CLAS:2020mom}, 
the internal structure of nucleons within nuclei \cite{Hen:2014nza}, 
neutrinoless double beta decay matrix elements \cite{Gallagher:2011zza}, 
and the properties of neutron stars \cite{Li:2018lpy}. 
This significance has spurred extensive theoretical and experimental efforts, employing kinematic conditions optimized to provide clear evidence of SRC 
\cite{Schiavilla:2006xx,Hen:2014nza,CLAS:2018yvt,CLAS:2019vsb,Kortelainen:2007rh,Li:2022fhh}. 

For nuclear many-body calculations, the usually adopted approach is 
to use unitary transformations to soften the interactions such that 
high-momentum correlations become sufficiently small that results converge within the tractable model spaces. Obviously, such transformations alter the SRCs dramatically such that their relation to the underlying NN interactions is lost. 
One of the most often used transformations is the similarity 
renormalization group (SRG) \cite{Bogner:2006pc,Anderson:2010aq}.
The impact of this transformation on the resulting nuclear wave functions 
is significant so that direct extractions of nuclear short-range
correlations (SRC) \cite{Tropiano:2021qgf,Hen:2016kwk} lead to results that are affected drastically by the transformation. This clearly shows the scheme dependence of the SRC that can only be tamed 
by restricting oneself to standard interactions (i.e. interactions that are driven by pion exchanges). 
For the SRG results, the expectation values that are required to extract 
observables are affected by the SRG transformation. It is therefore by now standard that corrections for these effects 
are applied to the operators required to extract such observables 
from the wave functions \cite{Schuster:2014lga,Schuster:2013sda,Neff:2015xda,
Gysbers:2019uyb,Tropiano:2020zwb,Friman-Gayer:2020vqn}.   

For the SRC, the SRG issue was in the past mostly avoided by applying 
Quantum Monte Carlo (QMC) methods that can be used directly 
with local phenomenological or chiral forces
\cite{Alvioli:2007zz,Feldmeier:2011qy,Hen:2016kwk,Chen:2016bde,Lynn:2019vwp,Cruz-Torres:2019fum,Piarulli:2022ulk} without SRG transformations. In this way, 
the one- and two-body nuclear densities were determined,    
NN SRC extracted and combined with the generalized contact formalism \cite{Weiss:2015mba} to explain related measurements
\cite{CLAS:2020mom,Weiss:2020bkp,Pybus:2020itv}. 
It was found that the ``nuclear contact'' ratio
are to a large extent independent of the type of interactions or 
to the regulator of the applied chiral forces \cite{Cruz-Torres:2019fum}.

Unlike other \textit{ab initio} methods, such as the no-core shell model (NCSM), in-medium SRG (IMSRG), or coupled-cluster (CC) method
\cite{Barrett:2013nh,
Hebeler:2015hla,Hergert:2015awm,
Carlson:2014vla,Hagen:2013nca},
QMC avoids SRG transformations, enabling direct SRC extraction. 
However, NCSM, IMSRG, and CC approaches can accommodate more 
interactions, e.g. using non-local 
regularization, and/or higher-order ChEFT contributions. 
The latter ones are essential for improving the accuracy of 
calculations in both light and medium-mass nuclei, for 
quantifying uncertainties arising from wave function dependencies, and also
for a better understanding of nuclear forces. 

However, for calculations based on these methods, unitary transformations, such as the SRG \cite{Bogner:2003wn,Roth:2010bm,Bogner:2006pc, Jurgenson:2009qs} are commonly employed to soften interactions, 
reducing high-momentum correlations to achieve convergence within the limited model spaces in nuclear many-body calculations.

It was argued that such correlations can 
be related to observables up to an uncertainty related 
to omitting, e.g., exchange currents and similar dynamical 
ingredients that prevent a one-to-one relation of the correlations 
and experimental data 
\cite{Furnstahl:2001xq,Bogner:2009bt,Furnstahl:2010wd}.
To address this, operator corrections
are routinely applied to recover high-momentum information \cite{Schuster:2014lga,Schuster:2013sda,Neff:2015xda,
Gysbers:2019uyb,Tropiano:2020zwb,Friman-Gayer:2020vqn}.


This work aims to investigate SRC
in light $s$- and $p$-shell nuclei using the Jacobi-NCSM (J-NCSM). 
To access sensible results, 
we perform the SRG transformation up to the two-nucleon 
cluster approximation on the wave functions, 
resulting in SRG-independent densities. 
Although this new 
approach is equivalent to the operator transformations
\cite{Schuster:2014lga,Schuster:2013sda,Neff:2015xda,
Gysbers:2019uyb,Tropiano:2020zwb,Friman-Gayer:2020vqn}, 
it is in practice easier to 
analyse electroweak processes since the nuclear 
structure related transformations are separated from the application of current operators. 
We provide the structure part conveniently in terms of transition densities \cite{Sun:2025yfo,Griesshammer:2024twu,Long:2025nll}. 
Based on these densities, we investigate short-range behavior and 
quantify the independence of SRC across different regularizations of chiral interactions. Thereby, we gain insight into the scheme 
dependence of the SRCs for a wider range of nuclear interactions. 

\textit{Model}---
The nuclear structure information is obtained from
one of the modern ${ab\ initio}$ methods, the 
J-NCSM with the Hamiltonian
\begin{equation}
H 
=\sum_{i<j}\frac{2}{A}\frac{\bm p_{ij}^2}{m} 
+\sum_{i<j=1}^A V_{ij}
+\sum_{i<j<k=1}^A V_{ijk} ,
\end{equation}
where $\bm p_{ij}$ is the pair momentum and $m$ is the nucleon mass.
For the NN and three-nucleon (3N)
forces $V_{ij}$ and $V_{ijk}$, the 
semilocal momentum-space-regularized (SMS)
NN potential at the order N$^{4}$LO$^{+}$ 
with momentum cutoffs $\Lambda_N=400$, $450$, $500$, and $550$~MeV \cite{Reinert:2017usi}
together with a consistently adjusted 3N potential at N$^{2}$LO 
(SMS N$^{4}$LO$^{+}$ + N$^{2}$LO) was used (see Table~1 of Ref.~\cite{Le:2023bfj}). These high precision 
chiral interactions yield an accurate description of the binding 
energies for light nuclei \cite{Maris:2020qne,LENPIC:2022cyu}.
The Schr\"odinger equation  is solved using the antisymmetrized harmonic oscillator (HO)
basis in  Jacobi coordinates with the HO frequency $\omega$
and model space truncated at a maximal HO excitation $N_\mathrm{HO}$ \cite{Liebig:2015kwa,Le:2020zdu,Le:2021gxa}.
In our calculations,
contributions of the NN potentials in partial waves up to
total NN angular momentum $J_{NN} = 6$ are included and for the $3N$ interaction 
all partial waves with total angular momentum $J_{3N} \le 9/2$ are taken into
account. 
Both the NN and 3N interactions are SRG-evolved on the 3N level to a flow parameter 
$\lambda$ of the order of $\lambda \approx 2$~fm$^{-1}$ to get converged energies.

To restore the high-momentum information, 
we calculate the unitary transformation corresponding to
the SRG evolution in the NN partial wave basis and get the SRG-evolved NN
relative wave functions in the Jacobi coordinate which are then used
to expand the nuclear many-body wave functions. 
The details of this
method is displayed in Section I of the supplement \cite{supp}.
This transformation is proven to be accurate 
for the case of the deuteron (see Section~II of~\cite{supp}). 
We have also compared our results for $^{4}$He with a Faddeev-Yakubovsky (FY) calculation \cite{Nogga:2000uu} based 
on bare interactions and found that our method can accurately reproduce the 
high-momentum part of two-body densities obtained from the wave functions. 
At the same time, the densities are almost independent on the SRG flow parameters.
The advantage of our method is that it allows us
to calculate SRG-independent expectation values of any two-body
operator, even for those including a momentum transfer or which do not conserve spin-isospin 
and isospin. Such densities have recently been employed for calculating nuclear radii \cite{Sun:2025yfo}.


\begin{figure*}[htb]
    \centering
    \includegraphics[width=0.9\linewidth]{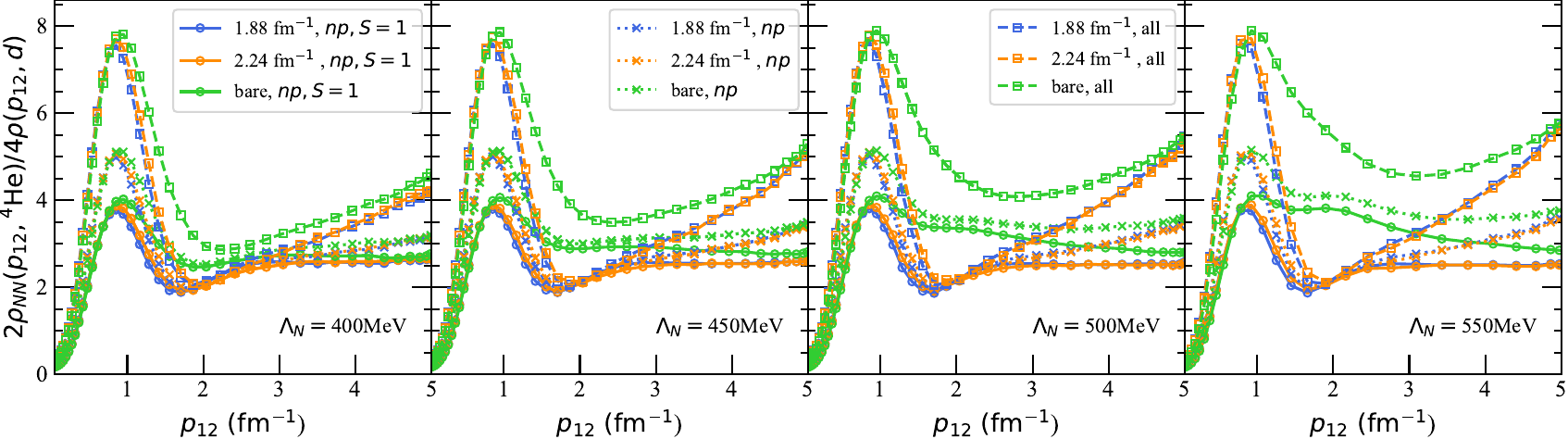}
    \includegraphics[width=0.9\linewidth]{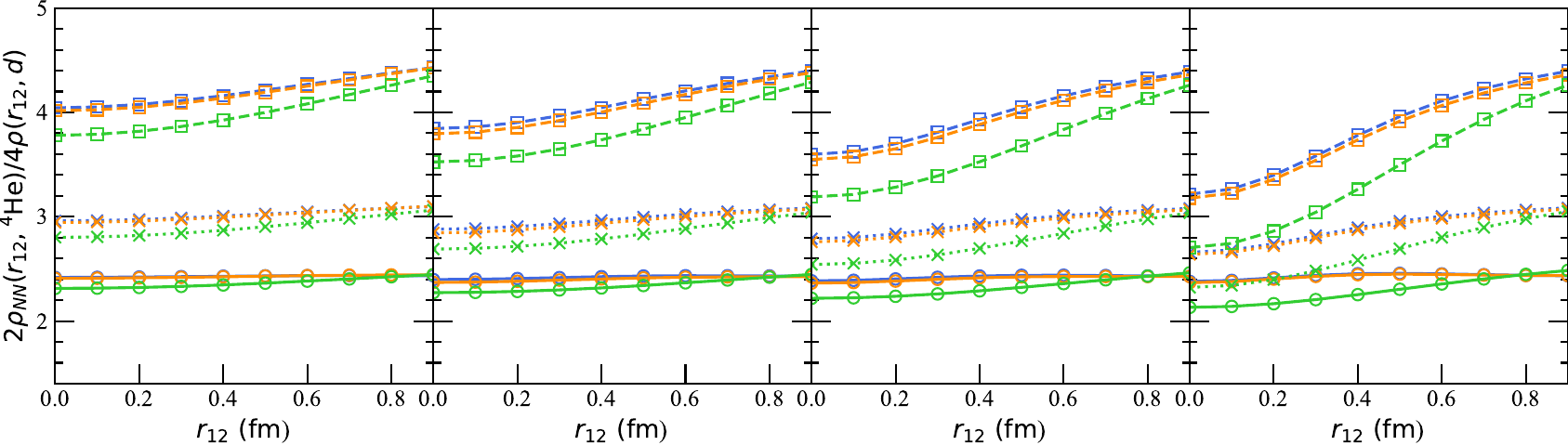}
    \caption{
    High-momentum and short-range behavior of the ratio 
    ${2\rho_{NN}(^{4}\mathrm{He})}/(4\rho(d))$ 
    for chiral forces
    SMS N$^{4}$LO$^{+}$ + N$^{2}$LO 
    with different momentum cutoffs
    ($\Lambda_N=400$, 450, 500, and 550 MeV).
    The two-body relative densities of $^{4}\mathrm{He}$ are calculated
    by J-NCSM model in a large HO model space $N_\mathrm{HO}=24$ with
    the interactions evolved with two flow parameters 1.88 and 2.24 fm$^{-1}$
    and by the FY approach with the bare interactions 
    (labeled as ``bare").
    The ratio between the $^{4}$He densities of 
    the $np$ with $S=1$ channel (circles),
    the $np$ channel (crosses), 
    and all channels (squares)
    and the deuteron density is presented.}
    \label{fig:a2-He4-d}
\end{figure*}

\textit{Results and discussions}---
Nuclear SRCs drive the high-momentum, short-distance structure of nucleon pairs. 
While the underlying two-body densities are sensitive to the specifics of chiral nuclear forces and reflect at short distances the model dependent repulsive core of the interactions, their ratios relative to a reference system like the deuteron ($d$) can reveal universal characteristics \cite{CiofidegliAtti:2015lcu,Weiss:2015mba,Carlson:2014vla}.
In Fig. \ref{fig:a2-He4-d}, we show
the ratio of the two-body relative densities between the $\alpha$
particle and the deuteron in the high-momentum and short-range parts.
The $^{4}$He densities are obtained by making the unitary transformation
on the wave function from the J-NCSM calculations with two SRG flow
parameters $\lambda=1.88\ \mathrm{fm}^{-1}$ and $2.24\ \mathrm{fm}^{-1}$
and the ``exact'' solution is obtained solving Faddeev-Yakubovsky (FY) equations. It is remarkable that the ratio is independent of the SRG parameter for all momenta including the very large ones. 
The ratio between $^{6}$He, $^{6}$Li and deuteron can be found in~\cite{supp}.
Our results reveal a striking universality: 
the density ratio in the spin $S=1$ neutron-proton ($np$) channel is independent of the chosen chiral interaction details and regularization scheme, 
for both high-momentum (up to $\approx 4.5~\text{fm}^{-1}$) and short-range components. 
This model independence, however, does not extend to the total $np$ channel or the sum of all two-body channels, which exhibit $\Lambda_N$ dependence in both J-NCSM and FY results. 

Looking in more detail at the $S=1$ ratios, ones  observe that there are some SRG parameter dependencies in the intermediate momentum region. These deviation gets systematically smaller for larger momentum but shows up as a mild SRG parameter dependence of the ratio at $r_{12}=0$. It is also interesting that the dependence of the SRG parameter is largest in a range of momenta for which the dependence on $\Lambda_N$, quantifying the dependence on the interaction, is largest. By performing FY calculations using SRG evolved interactions, it was checked that the dependence on $\lambda$ and $\Lambda_N$ are no numerical artifacts and indeed quantify the model dependence of the ratios of SRCs. 

We also checked a wider range of $\lambda$ values and confirmed a smooth 
transition from the bare results and the $\lambda = 1.88$ and $2.24$~fm$^{-1}$ results. This residual SRG dependence could in principle be reduced when transforming the densities on the three-nucleon level as it was done for the interactions. However, such calculations are difficult even using today's computing resources
and seem to be of less importance given that the variation of the ratios 
with respect to $\Lambda_N$ is of similar size as one can see when comparing the figures for $\Lambda_N=400$ and $550$~MeV.



\begin{figure}[htb]
    \centering
    \includegraphics[width=\linewidth]{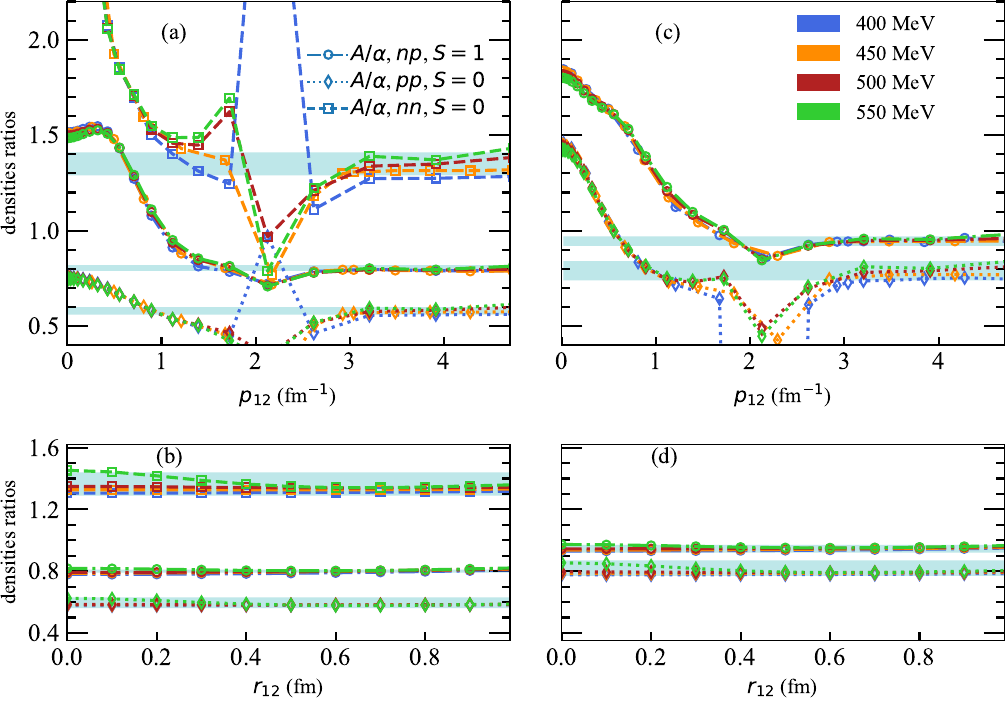}
    \caption{
    High-momentum and short-range behavior of the density ratio 
    in the $np$ with $S=1$ channel (unfilled circles),
    the $pp$ channel with $S=0$ (unfilled diamonds), 
    and the $nn$ channel with $S=0$ (unfilled squares)
    for chiral forces
    SMS N$^{4}$LO$^{+}$ + N$^{2}$LO 
    with different momentum cutoffs
    ($\Lambda_N=400$, 450, 500, and 550 MeV).
    $4\rho_{NN}(^{6}\mathrm{Li})/6\rho_{NN}(^{4}\mathrm{He})$ in $p$-space (a)
    and $r$-space (b).
    $4\rho_{NN}(^{6}\mathrm{He})/6\rho_{NN}(^{4}\mathrm{He})$ in $p$-space (c)
    and $r$-space (d). Note for $^{6}$Li, the results for $S=0$ $pp$ channel 
    are the same as those in $S=0$ $nn$ channel.
    The horizontal bands represent the range of extracted ratios for four chiral interactions. }
\label{fig:a2-He4}
\end{figure}

To further probe short-range correlations (SRCs), we examine density ratios relative to \({}^4\text{He}\) for various nucleon-nucleon channels in \({}^6\text{He}\) and \({}^6\text{Li}\) (Fig.~\ref{fig:a2-He4}). 
In such a way, one can investigate the density behaviors 
within the same theoretical and numerical method, 
and the two-body and three-body correlations are treated on the same footing using the J-NCSM calculations. 
The densities in
high-momentum and short-range
for different $\omega$ and $N_\mathrm{HO}$ tend to be
converged at a specific frequency, $\omega=16$ MeV in our
calculations and for $\omega=16$ MeV high momentum parts do not 
change as increasing $N_\mathrm{HO}$ (more details can be found in the supplements \cite{supp}).
Focusing on dominant SRC channels [$np$ ($S=1$), and $pp$/$nn$ ($S=0$)], we find that for the $np$ $S=1$ channel, the \({}^6\text{He}\)/\({}^4\text{He}\) and \({}^6\text{Li}\)/\({}^4\text{He}\) ratios are nearly invariant across different SMS interaction parametrizations at high momentum and short range. In the $pp$ $S=0$ channel, these ratios show only minor sensitivity to the interaction momentum cutoff $\Lambda_N$ (variations $\lesssim 0.05$), consistent with $\Lambda_N$-independence. Similarly, for the $nn$ $S=0$ channel in neutron-rich \({}^6\text{He}\), the ratio exhibits minimal $\Lambda_N$ dependence (e.g., varying from 1.3 to 1.4 at high momentum/short range), further supporting the scale independence.

The preceding analysis (Fig.~\ref{fig:a2-He4}) utilized a SRG
flow parameter $1.88~\text{fm}^{-1}$ and a single HO frequency $\omega=16$~MeV.
To test the SRG independence of our findings, calculations were repeated with $\lambda=2.24~\text{fm}^{-1}$ 
and for four values of $\omega$ around the optimal value of $\omega=16$~MeV. 
Fig.~\ref{fig:a2-A-He4-srg} shows momentum-space density ratios of $A=6$ nuclei (\({}^6\text{He}\), \({}^6\text{Li}\)) relative to \({}^4\text{He}\) in the $np$ $S=1$ channel, using both $\lambda$ values for SMS interactions with $\Lambda_N=450$ and $550$~MeV for $N_{\text{HO}}=12$.
The variations of the ratios is mostly seen in the medium-momentum regime ($1$--$3~\text{fm}^{-1}$). 
Remarkably, the dependence on $\lambda$ is less than 
on $\omega$. This indicates that the intermediate momentum 
regime is still not completely converged for $N_{\text{HO}}=12$
at $\omega \le 16$~MeV. 
The high-momentum behavior, however, is entirely insensitive to the SRG flow parameter and the HO frequencies. 
This insensitivity at high momentum and corresponding short 
range extends to other nucleon-nucleon channels. The results 
shows the robustness of the J-NCSM calculations for 
this quantity. Although the ratio is a property of the 
short-range or high momentum wave function, it is also remarkably independent of the cutoff $\Lambda_N$. 
 These results further substantiate the universality of the high-momentum components in appropriately constructed two-body relative density ratios across light nuclei.

\begin{figure}[htbp]
    \centering
    \includegraphics[width=\linewidth]{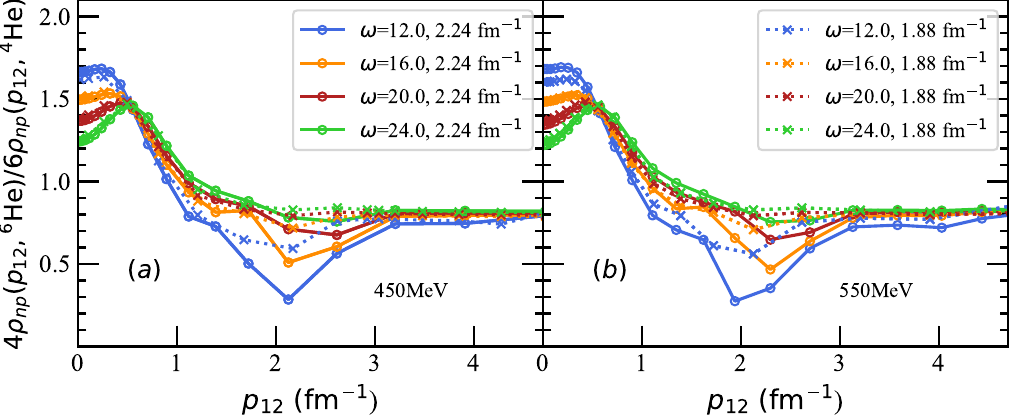}
    \includegraphics[width=\linewidth]{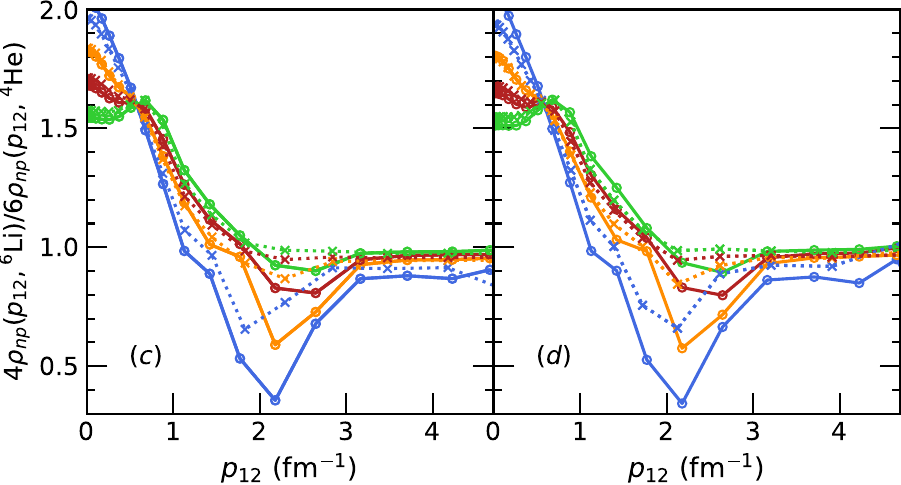}
    \caption{Flow parameter independence of the high-momentum
    and short-range behavior 
    for the $np$-chanal with $S=1$.
    $2\rho_{NN}({^{6}\mathrm{He}})/6\rho_{NN}(^{4}\mathrm{He})$ for 
    $\Lambda_N=450$ MeV (a) and $\Lambda_N=550$ MeV (b).
    $2\rho_{NN}({^{6}\mathrm{Li}})/6\rho_{NN}(^{4}\mathrm{He})$ for 
    $\Lambda_N=450$ MeV (c) and $\Lambda_N=550$ MeV (d).
    The ratio of four selected HO frequencies 12, 16, 20, and 24 are shown.
    Two flow parameters 1.88 fm$^{-1}$ (unfilled circles) and 2.24 fm$^{-1}$ (crosses) are adopted.}
\label{fig:a2-A-He4-srg}
\end{figure}

Several theoretical approaches can examine short-range scaling factors 
\cite{Feldmeier:2011qy,Rios:2013zqa,CiofidegliAtti:2015lcu,Weiss:2015mba,Bogner:2012zm,Hen:2016kwk,Chen:2016bde,Carlson:2014vla} and it can be extracted using density ratios relative to a reference nucleus---deuteron ($d$) or \({}^4\text{He}\) ($\alpha$)---in the limit of zero relative separation ($r_{12} \to 0$) or infinite relative momentum ($p_{12} \to \infty$) 
\begin{equation}
\begin{split}
\lim_{r_{12} \to 0}\frac{2\rho_{NN}(A,r_{12})}{A\rho_{NN}(d,r_{12})}
&\approx \lim_{p_{12} \to \infty}\frac{2\rho_{NN}(A,p_{12})}{A\rho_{NN}(d,p_{12})}; \\
\lim_{r_{12} \to 0}\frac{4\rho_{NN}(A,r_{12})}{A\rho_{NN}({}^4\text{He},r_{12})}
&\approx \lim_{p_{12} \to \infty}\frac{4\rho_{NN}(A,p_{12})}{A\rho_{NN}({}^4\text{He},p_{12})},
\end{split}
\label{eq:a2}
\end{equation}
 where the two-body relative density is normalized to the number of pairs of nucleons.
Figure~\ref{fig:a2} shows these $A$-dependent values for various spin-isospin channels, derived from densities of $ A\ge 4$ nuclei.
Crucially, these partial-wave scaling factors exhibit independence from the momentum cutoff $\Lambda_N$ of the SMS chiral interactions. 
This aligns with generalized contact formalism (GCF) predictions and indicates scale independence when compared with results from phenomenological or locally regulated chiral interactions \cite{Cruz-Torres:2019fum,Shang:2025vtd}. 
Observed smaller ratios for \({}^6\text{He}\) and \({}^6\text{Li}\) relative to \({}^4\text{He}\) underscore the influence of nuclear structure, such as $\alpha$-clustering with weakly bound valence nucleons, on these partial-wave SRC scaling factors.

Experimentally, scaling factors $a_2(A/d)$ are extracted from the ratios of cross-sections of $(e,e')$ scattering \cite{CLAS:2003eih}. For the usual nuclear interactions 
that are assuming a pion 
exchange long-range interaction phenomenologically or modern chiral interactions that systematically involve one- and 
more-pion exchanges, the ratios are rather independent 
of the interaction. Note that this is not the case for 
the short range wave functions themselves. So far, 
this has only been confirmed for local interactions 
since these observations are mostly based on Monte Carlo 
type calculations. Here, we extend this analysis to the 
non-local, high precision SMS interactions.
Our calculations show that for the total $NN$ density ratio: 
its momentum-space limit ($p_{12} \to \infty$) increases with $\Lambda_N$, while its coordinate-space limit ($r_{12} \to 0$) decreases. 
This dichotomy is linked to the characteristic center-of-mass momentum of SRC pairs \cite{CLAS:2018qpc}. 
Nevertheless, the $a_2({}^4\text{He}/d)$ values derived from our calculations in $r$-space with SMS interactions (varying from 3.2 to 4.0, see results shown in Fig. \ref{fig:a2-He4-d}) are consistent with the different experimental data summarized in Ref.~\cite{Hen:2012fm}, varying from $3.02\pm0.17$ to $3.80\pm0.34$. 
The analysis of inclusive measurements with generalized contact
formulism shows that the scaling factor is 
almost equal to the contact ratio of $np$ $S=1$ channel 
but underestimates the experimental data by about 20\%-30\% 
\cite{Weiss:2020bkp}.
This also holds for our calculations for these ratios of the $np$ $S=1$ channel.
Above all, achieving a better description of the scaling factor might consider another factor, 
such as 3N correlations or final-state interactions.

\begin{figure}[htbp]
\centering
\includegraphics[width=0.9\linewidth]{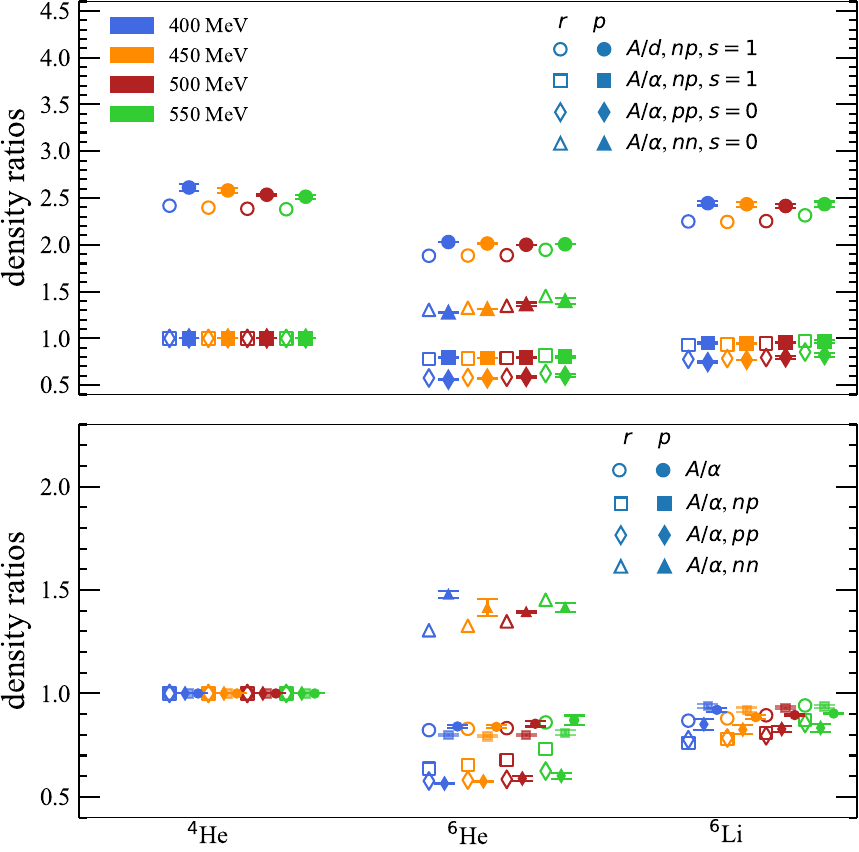}
\caption{Density ratios for different spin-isospin channels for chiral forces
SMS N$^{4}$LO$^{+}$ + N$^{2}$LO 
with different momentum cutoffs ($\Lambda_N=400$, 450, 500, and 550 MeV).
Upper panel: Ratios between $^{4,6}$He, $^{6}$Li and $d$ for the spin $s=1$ $np$ channel 
in the coordinate space ($r$-space, circles) and momentum space ($p$-space, filled circles).
Ratios between $^{4,6}$He, $^{6}$Li and $^{4}\mathrm{He}$ ($\alpha$) for the spin $s=1$ $np$ channel (squares),
$s=0$ $pp$ channels (diamonds), and $s=0$ $nn$ channels (triangles)
in the $r$-space (nonfilled symbols) and $p$-space (filled symbols).
Bottom panel: Ratios with respect to $\alpha$ 
partial but for
all two-body channels (circles), 
the $np$ channel (squares),
$pp$ channels (diamonds), 
and $nn$ channels (triangles)
in the $r$-space (nonfilled symbols) and $p$-space (filled symbols).
}
\label{fig:a2}
\end{figure}

We also analyze density ratios with \({}^4\text{He}\) as the reference nucleus. For dominant two-body channels [$np$ ($S=1$) and $pp/nn$ ($S=0$)], these ratios remain independent of the interaction's momentum cutoff $\Lambda_N$, see upper panel of Fig. \ref{fig:a2}. 
Including contributions from $np$ ($S=0$) and $pp/nn$ ($S=1$) channels introduces a slight $\Lambda_N$ dependence, 
though variations across the four SMS interactions tested do not exceed 10\%
(see the bottom panel of Fig. \ref{fig:a2}). 
Remarkably, when considering ratios derived from total two-body densities (summed over all channels) relative to \({}^4\text{He}\), 
we observe almost complete insensitivity to the interaction details (including $\Lambda_N$) in both momentum and coordinate space. 
This starkly contrasts with the $\Lambda_N$ dependence found for total density ratios referenced to the deuteron. 
Such resilience to interaction details concerning $\alpha$ particle might be due to $^6$Li ($^{6}$He) can be regarded as the combination of an innerest core -- $\alpha$ particle with a weakly bound $d$ ($2n$), thus the high-moment correlation mainly reflected by the compact core with low center of mass momentum. 
\textit{Summary}---
We introduce a novel method to construct high-resolution nuclear wave
functions from J-NCSM solutions obtained with SRG-evolved SMS N\(^4\)LO\(^+\) + N\(^2\)LO chiral interactions. 
This approach 
can provide SRG flow-parameter-independent nuclear densities, 
which allows us to study short-range behaviors and to check their dependence 
on the inherent regularization of the chiral interactions.
We investigated the independence of two-nucleon short-range behaviors on the inherent momentum cut-offs of the chiral interactions and found
that for the $np$ $S=1$ channel, when referenced to the deuteron, the conclusion aligns with QMC and the 
local interactions studied in these works.
Additionally, density ratios relative to the $\alpha$ particle demonstrate robust independence from the specific details of the chiral interactions. 
This work advances the understanding of short-range universality in the context of the chiral forces and provides a versatile framework that can be extended to other \textit{ab initio} nuclear theories that rely on 
SRG evolution as IMSRG and CC and which allow the 
extension to heavier nuclei.

\begin{acknowledgments}
This work was supported in part by the European
Research Council (ERC) under the European Union's Horizon 2020 research
and innovation programme (grant agreement No. 101018170),
by the MKW NRW under the funding code NW21-024-A,
and by the CAS President's International Fellowship Initiative (PIFI) under Grant No.~2025PD0022.
 The numerical calculations were performed on JURECA
of the J\"ulich Supercomputing Centre, J\"ulich, Germany.
\end{acknowledgments}


\putbib

\end{bibunit}

\clearpage 

\setcounter{figure}{0} 
\setcounter{table}{0} 

\begin{bibunit}



\title{Supplement for ``Universality of nucleon short-range behavior with chiral forces" }

\author{Xiang-Xiang Sun\orcidlink{0000-0003-2809-4638}}
\affiliation{Institute for Advanced Simulation (IAS-4), Forschungszentrum J\"{u}lich, D-52425 J\"{u}lich, Germany}

\author{Hoai Le\orcidlink{0000-0003-1776-9468}}
\affiliation{Institute for Advanced Simulation (IAS-4), Forschungszentrum J\"{u}lich, D-52425 J\"{u}lich, Germany}

\author{Ulf-G. Mei{\ss}ner\orcidlink{0000-0003-1254-442X}}
\affiliation{Helmholtz-Institut~f\"{u}r~Strahlen-~und~Kernphysik~and~Bethe~Center~for~Theoretical~Physics, Universit\"{a}t~Bonn,~D-53115~Bonn,~Germany} 
\affiliation{Institute for Advanced Simulation (IAS-4), Forschungszentrum J\"{u}lich, D-52425 J\"{u}lich, Germany}
\affiliation{Peng Huanwu Collaborative Center for Research and Education, International Institute for Interdisciplinary and Frontiers, Beihang University, Beijing 100191, China}
\affiliation{CASA, Forschungszentrum J\"{u}lich, 52425 Ju\"{u}lich, Germany}

\author{Andreas Nogga\orcidlink{0000-0003-2156-748X}}
\affiliation{Institute for Advanced Simulation (IAS-4), Forschungszentrum J\"{u}lich, D-52425 J\"{u}lich, Germany}
\affiliation{CASA, Forschungszentrum J\"{u}lich, 52425 Ju\"{u}lich, Germany}

\date{\today}
\maketitle

\onecolumngrid

\makeatletter
\renewcommand{\c@secnumdepth}{0}
\makeatother

\section{SRG evolution of J-NCSM wave functions}
\label{A1}
In most of the state-of-the-art nuclear 
\textit{ab initio} calculations, 
to get convergent energy and its corresponding eigenstates 
$|\Psi_\lambda\rangle$ 
the adopted realistic nuclear Hamiltonian 
is usually softened 
via a unitary transformation, 
for example, the most commonly used
SRG transformation,  
$H_\lambda = U_\lambda H_0 U^{\dagger}_\lambda 
= T_\mathrm{rel} + V_\lambda$ with the bare interaction Hamiltonian $H_0$ and
the relative kinetic energy $T_\mathrm{rel}$. 
$V_\lambda$ and unitary transformation operators 
are governed by the SRG flow equations 
\begin{equation}
\frac{dV_\lambda}{d\lambda} = [\eta_\lambda,H_\lambda], \quad
\frac{dU_\lambda}{d\lambda} = U_\lambda\eta_\lambda,
\label{eq:srg}
\end{equation}
where $\eta_\lambda=[T_\mathrm{rel},H_\lambda]$ 
and 
$\lambda$ is the flow parameter. This notation is the same as in Ref.~\cite{Sun:2025yfo}.

The SRG-cutoff-independent expectation value of any 
``bare" operator $O$ can be calculated by
\begin{equation}
\langle O \rangle  \equiv  
\langle \Psi_{\lambda=0} | O | \Psi_{\lambda=0} \rangle 
=\langle \Psi_{\lambda} | O_\lambda | \Psi_\lambda \rangle
=\langle \Psi_{\lambda} | U_\lambda O U^{\dagger}_\lambda | \Psi_\lambda \rangle,
\end{equation}
where $|\Psi_\lambda \rangle = U_{\lambda}|\Psi_{\lambda=0} \rangle$
with the bare interaction solution $|\Psi_{\lambda=0} \rangle$  and 
$O_\lambda=U_\lambda O U^{\dagger}_\lambda$ . 
This can usually be achieved by constructing 
the matrix elements of $U$ using energy eigenstates \cite{Anderson:2010aq}
\begin{equation}
U_\lambda = \sum_{i} |\Psi_{\lambda}^{i}\rangle
\langle \Psi_{\lambda=0}^{i}|,
\end{equation}
or 
evolving the operator via the same flow equations.  
The SRG cutoff independent expectation value
can be directly obtained by constructing the nuclear many-body wave function 
corresponding to the bare interaction $|\Psi_{\lambda=0}\rangle$
based on the solution of low-resolution interactions $V_\lambda$, i.e.,
calculating $U^{\dagger}_\lambda|\Psi_{\lambda}\rangle$. We truncate 
the unitary transformation at a two-body level, which has been applied in several studies on nucleon short-range correlation
\cite{Tropiano:2021qgf,Neff:2015xda}
and apply this method to finite nuclei using the wave functions 
of low-resolution interactions from the Jacobi-NCSM approach
\cite{Liebig:2015kwa,Le:2020zdu}, which is performed in momentum space.

The $A$-body wave function can be expanded by a harmonic oscillator (HO) basis and is composed of different blocks labeled as $|\Psi^{N_\mathrm{HO}JTM_T}_\lambda\rangle$ where $N_\mathrm{HO}$ is the total HO quantum number of the $A$-body system and $J$, $T$, and $M_T$ for the 
angular momentum, isospin, and isospin projection of the nucleus, respectively. For our aim, it is most convenient to 
represent the wave function in a basis that couples out one nucleon pair. 
This can be constructed by 
combining the anti-symmetrized wave functions of 
$(A-2)$-subsystem and two-body subsystem
in two-body relative HO basis in J-NCSM model
\begin{align}
\begin{split}
|\Psi^{N_\mathrm{HO}JTM_T}_\lambda \rangle  
= &  
\sum_{\gamma_{A-2},\alpha} F_{\lambda}^{\alpha,\gamma_{A-2}} 
C^{TM_T}_{t_{12}T_{A-2}; m_t M_T-{m_t}_{12}} 
C^{JM}_{j_{12}I, m_{12}M-m_{12}} \\
&\times |n_{12}(l_{12}s_{12})j_{12} {m_j}_{12}; t_{12} {m_t}_{12}\rangle 
|\gamma_{A-2}: N_{A-2} IM-{m_j}_{12}; T_{A-2} M_T-{m_t}_{12} \rangle,
\end{split}
\label{wf0}
\end{align}
where $\alpha$ refers to all the quantum numbers of
two-body relative states
$\alpha \equiv (n_{12}(l_{12}s_{12})j_{12} {m_j}_{12}; t_{12} {m_t}_{12})$, 
$\gamma_{A-2}$ for $(A-2)$-body cluster states 
and the relative motion between them. $I$ is the spin for $(A-2)$-subsystem, its third component is $M-m_{j_{12}}$, and $N_{(A-2)}$ is the HO quanta. 
The iso-spin and its projection of $(A-2)$-subsystem are $T_{(A-2)}$ and $M_T-m_{t_{12}}$.
The two parts are coupled using Clebsch-Gordan coefficients labelled as $C$. The two body state is defined by its HO quantum number $n_{12}$, the orbital 
angular momentum $l_{12}$, the spin $s_{12}$ and the total two-nucleon angular momentum $j_{12}$ and its third component ${m_j}_{12}$. $t_{12}$ labels the isospin of the nucleon pair and ${m_{t}}_{12}$ its third component.

The coefficient $F_{\lambda}^{\alpha,\gamma_{A-2}}$ 
is obtained from the diagonalization of 
the Hamiltonian matrix with softened interaction and certainly 
depends on the SRG flow parameters. 
Since limiting unitary transformation operators up to the two-body level, 
the high-resolution wave function
$|\Psi^{N_\mathrm{HO}JTM_T}\rangle=U^{\dagger}_\lambda|\Psi^{N_\mathrm{HO}JTM_T}_\lambda\rangle$, 
can be obtained by calculating
$U^{\dagger}_\lambda|n_{12}(l_{12}s_{12})j_{12}t_{12} {m_t}_{12}\rangle$, 
which can be achieved by solving
the flow equations of $U^{\dagger}_\lambda$. 

The SRG evolution of the two-body NN forces 
can be directly achieved in the two-body partial wave basis 
$|p(l_{12}s_{12})j_{12}t_{12}{m_{t}}_{12}\rangle$. 
$U^{\dagger}_\lambda|n_{12}(l_{12}s_{12})j_{12}t_{12} {m_t}_{12}\rangle$ 
can also be solved in a similar way. 
Let us define
\begin{equation}
\phi^{j_{12}s_{12}}_{n_{12},l_{12}l'_{12}}(p;\lambda) \equiv 
\langle n_{12}(l_{12}s_{12})j_{12}|U^{\dagger}_{\lambda}
|p(l'_{12}s_{12})j_{12}\rangle.
\end{equation}
Its solution is governed by the following flow equations 
\begin{equation}
\frac{d}{d\lambda} 
 \phi^{j_{12}s_{12}}_{n_{12},l_{12}l'_{12}}(p;\lambda)  = 
-\sum_{\tilde{l}}\int dp'{p'}^{2}
\phi^{j_{12}s_{12}}_{n_{12},l_{12}\tilde l}(p';\lambda)
(\frac{p'^{2}}{2\mu}-\frac{p^{2}}{2\mu})
V_{\tilde{l}{l'_{12}}}^{j_{12}s_{12}}(p{'},p;\lambda),
\end{equation}
where $V_{\tilde{l}{l}}^{j_{12}s_{12}}(p^{'},p;\lambda)$ 
is the SRG evolved interaction in momentum space. 
Taking HO wave functions in momentum space as the initial conditions, 
this equation can be solved simultaneously 
with the flow equation of interactions. 
We can calculate 
$U^{\dagger}_\lambda|
n_{12}(l_{12}s_{12})j_{12}t_{12} {m_t}_{12}\rangle$ 
in the partial wave basis as
\begin{equation}
\label{eq:hoevolvedwf}
\tilde{R}^{j_{12}s_{12}}_{{n_{12},ll_{12}}}(p;\lambda)
=
\langle p(ls_{12})j_{12}|
U_{\lambda}^{\dagger}
|n_{12}(l_{12}s_{12})j_{12} \rangle 
=\sum_{n'}R_{n'l}(p)\int dp'p'^{2}
\phi_{n',ll_{12}}^{j_{12}s_{12}}(p';\lambda)
 R_{n_{12}l_{12}}(p'),
\end{equation}
where $R_{nl}(p) =\langle p(ls)j|n(ls)j\rangle$ 
is the HO wave function in momentum space. We stress that this procedure only involves the two-body HO basis.
Finally, the high-resolution many-body wave functions 
can be directly obtained using this SRG evolved two-body partial wave basis and the expansion of Eq.~\eqref{wf0}.

\section{Benchmark calculations for deuteron wave functions}
\label{sec:A2}
\begin{figure}[tbp]
\centering
\includegraphics[width=0.85\linewidth]{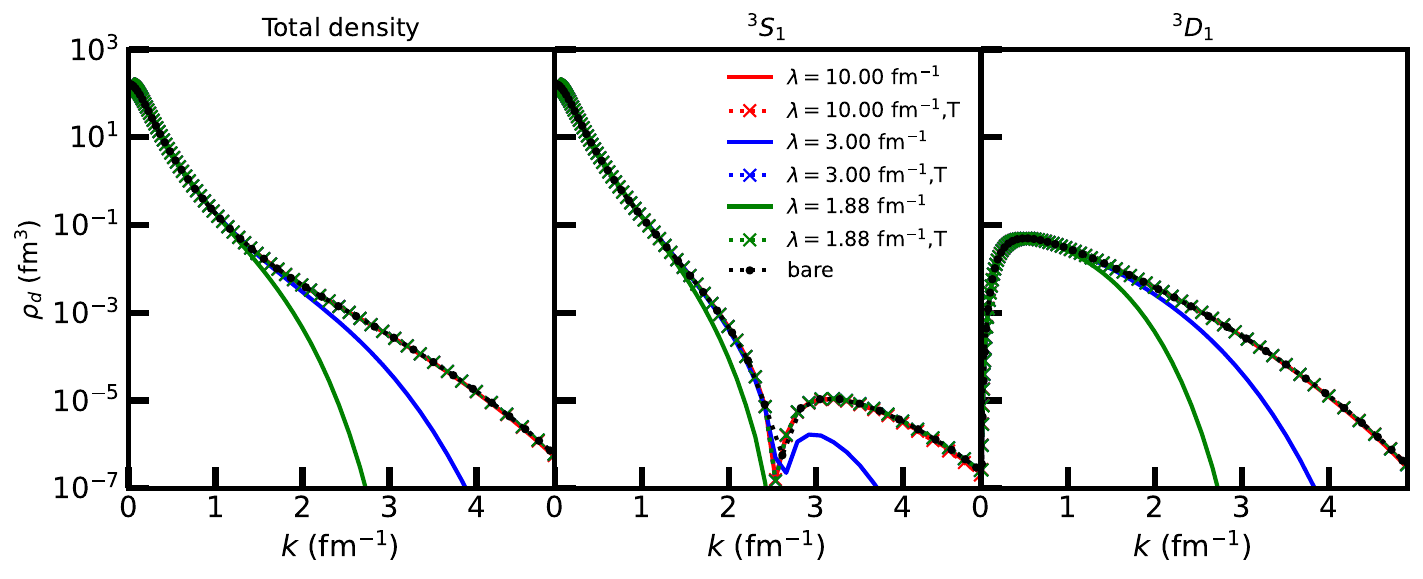}
\caption{Deuteron density distributions in momentum space using the SMS N$^{4}$LO$^{+}$ 
interaction with a cutoff of $450$~MeV. 
All results have been obtained using a HO basis. Solid lines are results using directly the wave functions obtained solving the Schrödinger equation for SRG-evolved interactions. Dotted lines and crosses are obtained by using the transformed wave functions of Eq.~(\ref{eq:hoevolvedwf}). The results for the bare interaction is shown as a black dotted line with circles.
The total deuteron densities is normalized to $\int_0^{\infty} dk k^2 \, \rho(k) = 1 $.}
\label{fig:deu}
\end{figure}

Deuteron wave functions can be exactly 
obtained by using the two-body relative HO basis, which is an ideal system for checking the 
above-mentioned method. 
By using the state-of-the-art
chiral NN interaction, namely the semilocal momentum-space-regularized (SMS)
NN potential at the order N$^{4}$LO$^{+}$ with momentum cutoff $\Lambda_N=$ 450 MeV
evolved to three SRG cutoffs $\lambda=$ 1.88 fm$^{-1}$, 3.00 fm$^{-1}$, and 10.00 fm$^{-1}$,
the deuteron wave functions are obtained. 
The solutions calculated with the bare interaction are also shown for comparison. 
In Fig. \ref{fig:deu}, 
we show the total density and the contribution from ${}^{3}{S}^{}_{1}$ and
${}^{3}{D}^{}_{1}$ channels as a function of relative momentum.
It is clear that both the total density and its partial wave components 
in the high momentum region
strongly depend on the SRG cutoff when using low-resolution interactions
because the tensor couplings between $S$- and $D$-waves are suppressed by
the unitary transformation.
When applying the SRG evolved two-body relative wave functions 
$\tilde{R}^{j_{12}s_{12}}_{{n_{12},l'_{12}l_{12}}}(p; \lambda)$ to 
the deutron wave functions, 
our calculations show that the strong dependence of
total density and also its ${}^{3}{S}^{}_{1}$ and
${}^{3}{D}^{}_{1}$ components on the SRG cutoffs all disappear,
meaning the method as mentioned earlier is exact not only 
for deuteron density but also wave functions.

\section{Two-body relative densities of \texorpdfstring{$^{4,6}{\rm He}$}{4,6-He} and \texorpdfstring{$^{6}{\rm Li}$}{6-Li}}
In Section \ref{sec:A2} of the supplement, we have shown that the wave function back transformation for the HO basis is accurate resulting in a SRG parameter independent density that agrees with the bare solution.
In this part, we check the accuracy of the $A=4$ system using the Jacobi-NCSM approach \cite{Liebig:2015kwa}. We can not expect exact independence of the wave function/densities from the SRG parameter using Eq.~\eqref{eq:hoevolvedwf} anymore because of the restriction to only implement the back transformation on the two-body level. 
We will however see below that the independence can be achieved with high accuracy even in this approximation.

To this aim, we use the wave function 
from J-NCSM
calculations corresponding to the low-resolution interactions
to construct high-resolution wave functions for light nuclei by expanding in the transformed HO wave functions Eq.~\eqref{eq:hoevolvedwf}. 
The momentum space densities are defined by 
\begin{equation}
\rho_{NN}(p) = \frac{A(A-1)}{2}\frac{1}{(2J+1)}\sum_{M}\langle \Psi^{JMM_T}|\frac{\delta(p-p_{12})}{p^{2}_{12}}|\Psi^{JMM_T}\rangle.
\label{eq:den_p}
\end{equation}
Note that the two-body density is normalized to the number of nucleon pairs $A(A-1)/2$. 
The configuration space densities can be straightforwardly obtained using Fourier transforms of the  wave functions of Eq.~\eqref{eq:hoevolvedwf} or of the HO 
wave functions.

\begin{figure}[tbp]
\centering
\includegraphics[width=.8\linewidth]{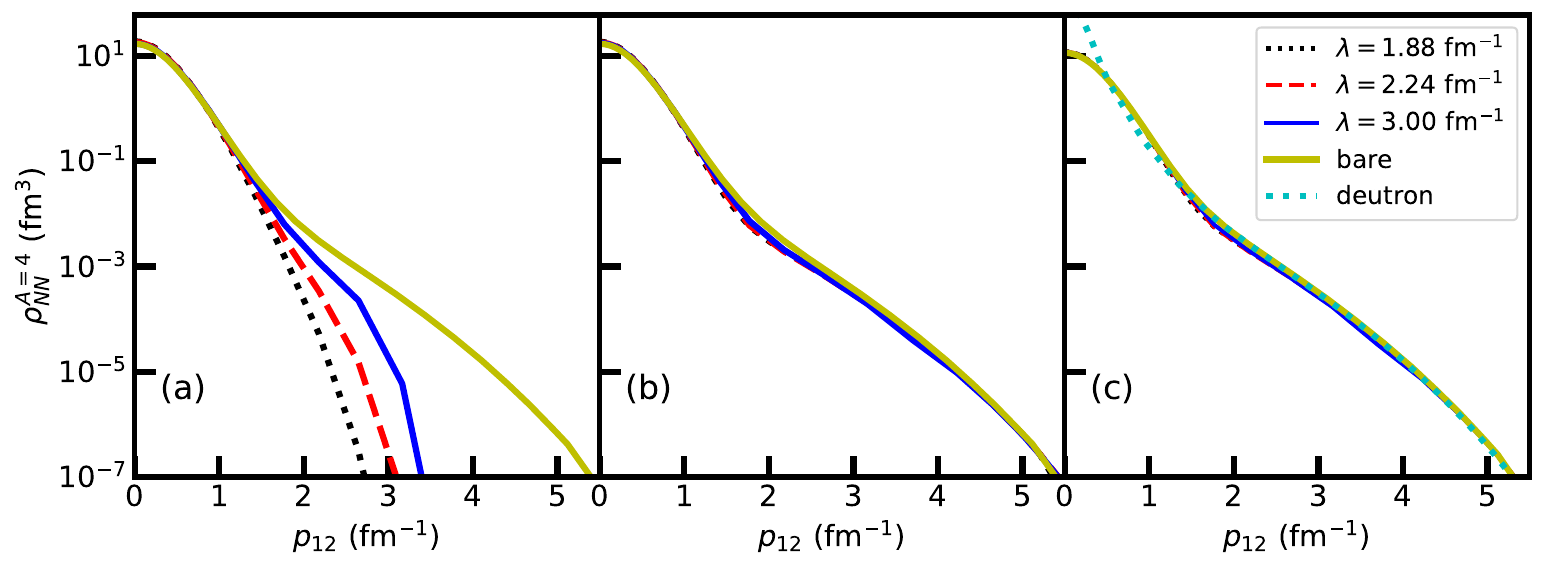}
\caption{Two-body density distributions of $^{4}$He in momentum space by using 
two-body HO functions (a) and evolved two-body relative wave functions (b), and
the contribution from $np$ channel to the two-body density using our method (c).
The adopted interaction is SMS N$^{4}$LO$^{+}$ (450 MeV) + N$^{2}$LO 
evolved to three different SRG parameters.
The two-body density calculated from FY equations using bare interaction (labeled as `bare')
and the deuteron density is also given for comparison.}
\label{fig:he4-den1}
\end{figure}

As an example, we employ 
J-NCSM calculations with NN interaction  
SMS N$^{4}$LO$^{+}$ ($\Lambda_N=450$ MeV) 
and three-nucleon force (3NF) in N$^{2}$LO 
(for more details about the parameterization of interactions, 
see Ref. \cite{Reinert:2017usi,Maris:2020qne,Le:2023bfj}) 
evolved to different SRG cutoffs (1.88, 2.24, and 3.00 fm$^{-1}$). The calculations use a
HO frequency $\omega=16$~MeV and the total HO energy quantum number
of $N_{\rm HO}^{}=20$. With this basis size, 
the calculations of binding
energy is well converged and independent of the HO frequencies around the optimal value of $\omega=16$~MeV. 
The two-body density distributions 
in momentum space calculated by using
low-resolution wave function expanded 
in two-body HO basis are shown in
Fig. \ref{fig:he4-den1}~(a). 
It is clear that the two-body density 
strongly depends on the SRG parameter 
and with the increase of this value,
more high-momentum correlations are included 
such that the high momentum tail appears gradually. 
Full agreement with the bare result can however not be 
achieved because the J-NCSM does not anymore converge for larger SRG parameters. 
After using the SRG-evolved two-body relative function 
from Eq.~\eqref{eq:hoevolvedwf}, 
the dependence on the SRG parameter mostly disappears
as shown in Fig. \ref{fig:he4-den1}~(b). It is reassuring that the result also agrees  with   
Faddeev-Yakubovsky (FY) 
calculations \cite{Nogga:2000uu} using the bare interactions.
This shows that the high-resolution densities can be 
successfully obtained by using low-resolution 
wave functions expanded by the SRG-evolved 
two-body relative functions.
There are only very small differences for $1.5<p_{12}<2.8\  \mathrm{fm}^{-1}$, 
which are either due to truncating the unitary transformation of Eq.~\eqref{eq:hoevolvedwf} to the two-body level and/or due to   
the absence of the induced forces at the four-nucleon level in the J-NCSM approach.
In Ref. \cite{Neff:2015xda}, by evolving the density matrix operators with the 
flow equations, the calculations of $^{4}$He two-body densities have shown similar
conclusions. Also their results have a weak dependence on SRG parameters in
the medium-momentum part. These appear to be slightly more pronounced than in our calculations, which might be due to the absence of induced three-body forces or because the bare interactions are different to ours.

\begin{figure}[tbp]
    \centering
    \includegraphics[width=.85\linewidth]{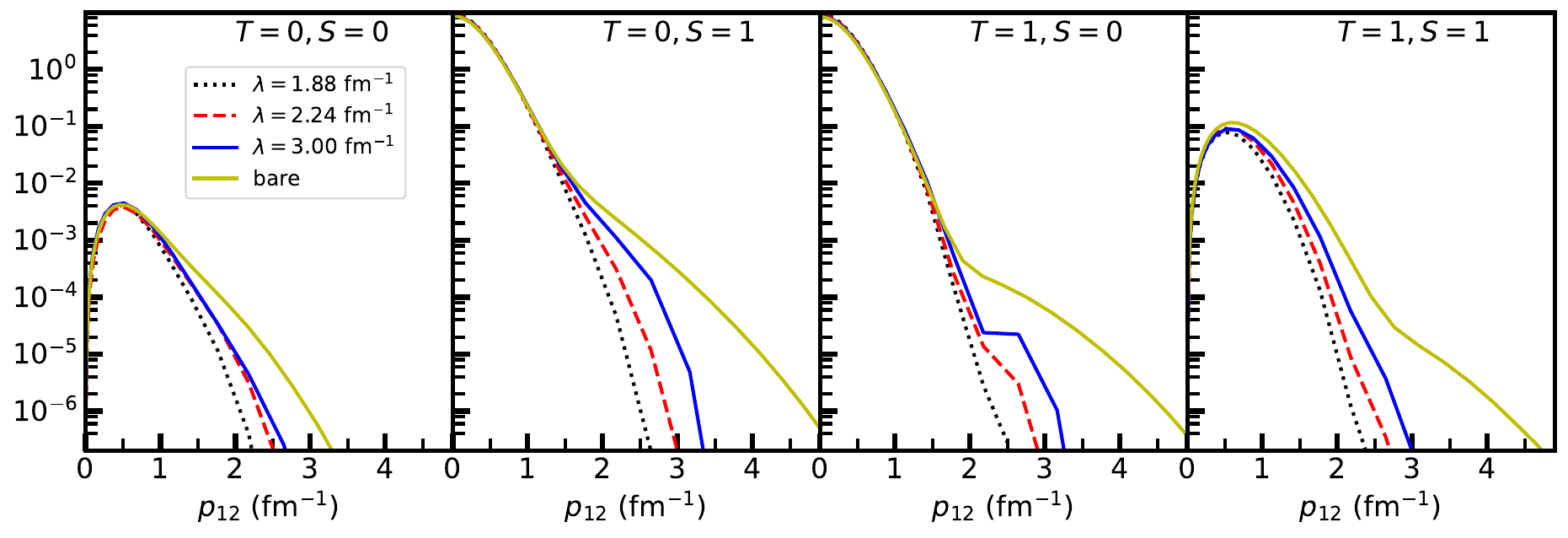}
    \includegraphics[width=.85\linewidth]{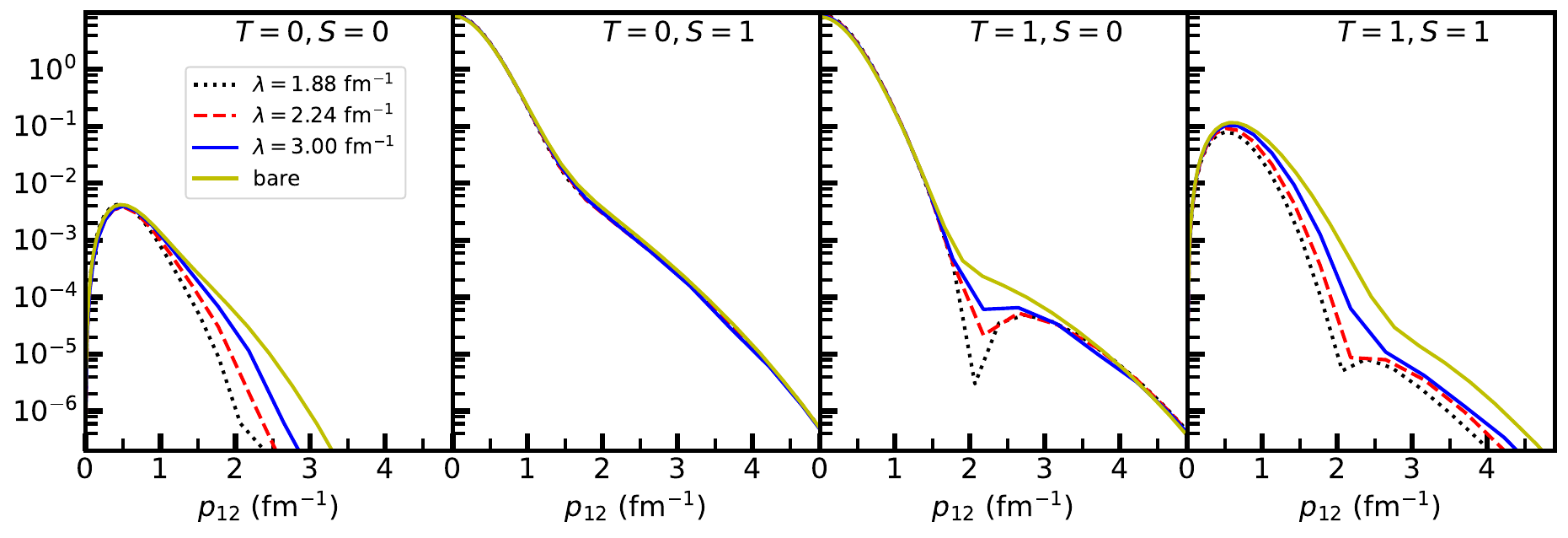}
    \caption{Two-body density distributions in different spin ($S$)-isospin ($T$) channel of $^{4}$He 
    in momentum space by using the two-body relative HO basis (the upper panel) and
    evolved two-body relative wave functions (the bottom panel).
    The adopted interaction is SMS N$^{4}$LO$^{+}$ (450 MeV) + N$^{2}$LO 
    evolved to three different SRG parameters $1.88$, $2.24$, and $3.00$~fm$^{-1}$.
    The results from FY with the bare interactions, 
    labeled as `bare', are also given for comparison.}
    \label{fig:he4-den-st}
\end{figure}

To get insights into the two-body densities
the densities are decomposed into different isospin($T$)-spin($S$) channels
($T=0,1$ and $S=0,1$) shown in Fig. \ref{fig:he4-den-st} and compared with
the results from the FY method with the bare interactions.  
Interestingly, for the $(T,S)=(0,1)$ channel, the deuteron channel, which is strongly
influenced by the tensor force, the results of different SRG parameters are well consistent with the bare interaction results. Apparently, truncating the unitary transformation on the two-body level works extraordinarily well for this case. 

The contribution of $(T,S)=(0,0)$ and $(1,1)$ to the total density is small. But the approximations involved 
the calculation show up by a visible dependence on the SRG parameter and some deviation from the bare result. 
This possibly affects observables that enhance the contribution of this part of the wave functions. 

The second important contribution to the wave function is given by the $(T,S)=(1,0)$ channel. 
In this case, the high-momentum and low-momentum part  is well restored by the two-body
level unitary transformation. For intermediate momenta 
around 2 fm$^{-1}$, there is still a residual dependence 
on the SRG parameter. This dependence can also be seen in the total density, see Fig.~\ref{fig:he4-den1}~(b),
but, fortunately, is less pronounced because the 
$(T,S)=(0,1)$ channel is more important in this momentum 
region. 

Again, similar observations were done in 
Ref.~\cite{Neff:2015xda} which supports the 
equivalence of evolving the operators and the densities. 
We emphasize that our transformed densities can be directly 
applied to very different operators so that calculations 
of various observables are simplified in this method.

\begin{figure}[tbp]
\centering
\includegraphics[width=\linewidth]{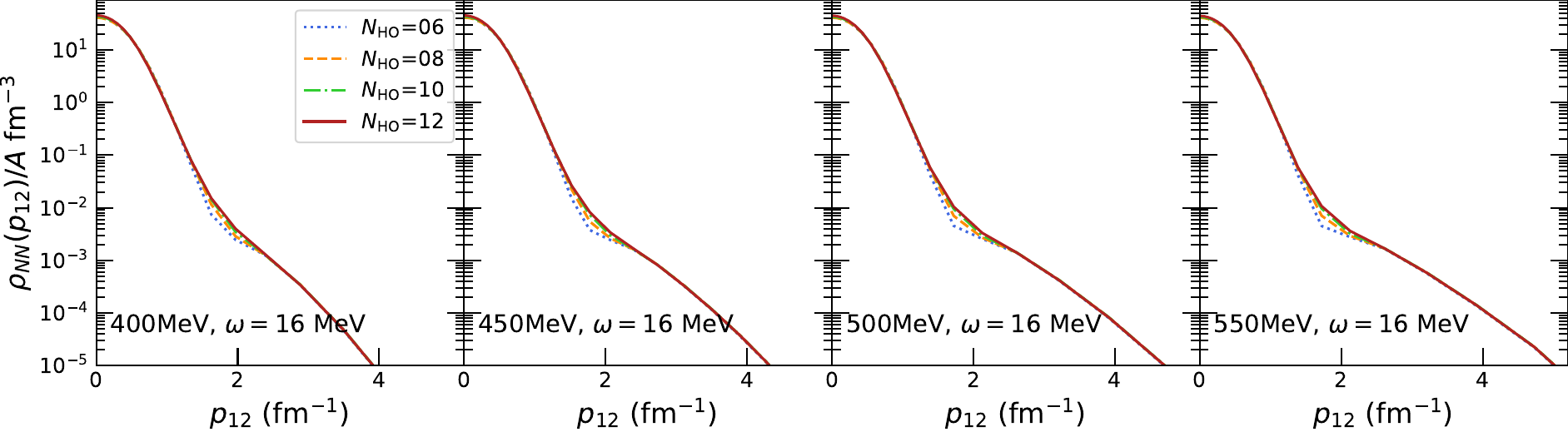}
\includegraphics[width=\linewidth]{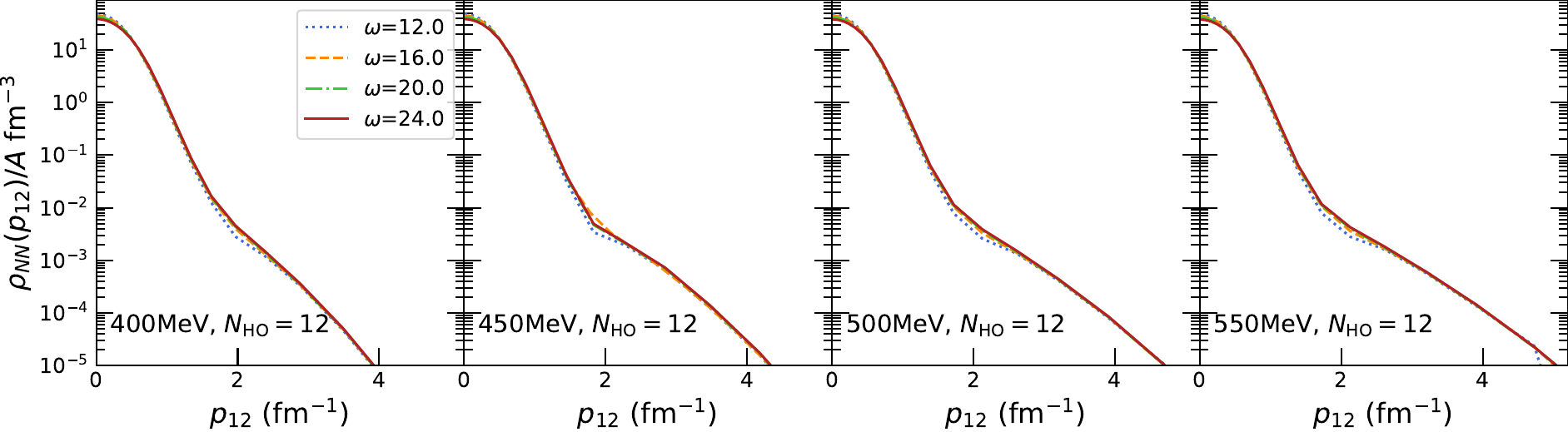}
\caption{Two-body density distributions of $^{6}$Li in momentum space from J-NCSM
calculations using the chiral forces N$^{4}$LO$^{+}$ + N$^{2}$LO with the momentum cut-off
$\Lambda_N=400$, 450, 500, and 550 MeV.
The upper panel, densities are calculated with 
the fixed HO frequency 16 MeV for $N_\mathrm{HO}=6$, 8, 10, and 12.
The bottom panel, densities are calculated with 
the fixed HO basis space $N_\mathrm{HO}=12$ for four HO frequencies 12 MeV, 16 MeV, 20 MeV, and 24 MeV.}
\label{fig:Li6-denp}
\end{figure}

\begin{figure}[tbp]
\centering
\includegraphics[width=\linewidth]{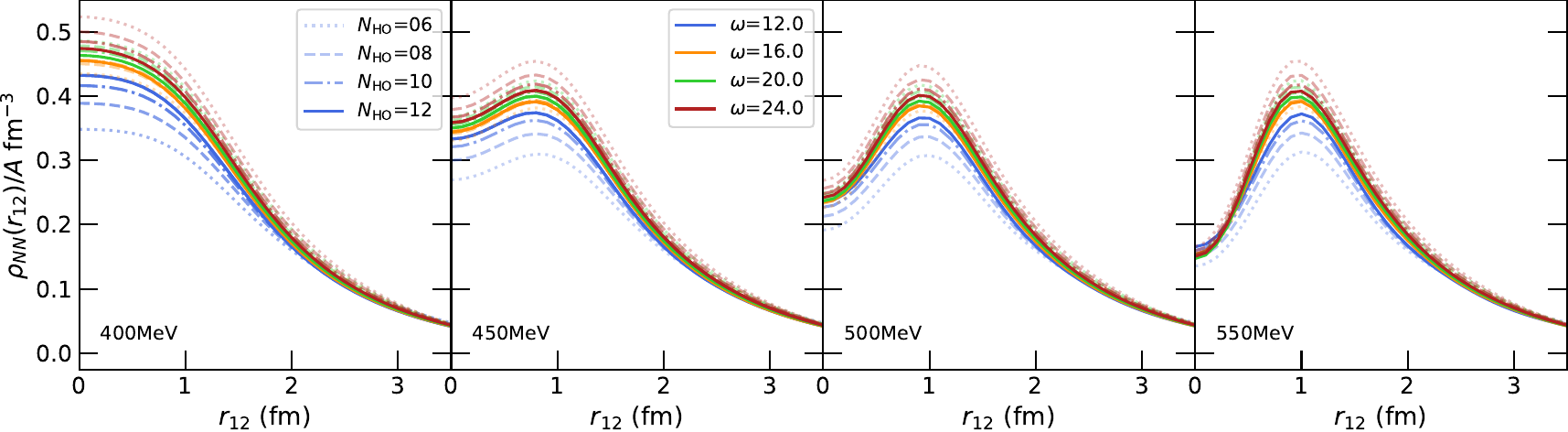}
\caption{Two-body density distributions of $^{6}$Li in $r$-space from J-NCSM
calculations using the chiral forces N$^{4}$LO$^{+}$ + N$^{2}$LO with the momentum cut-off
$\Lambda_N=400$, 450, 500, and 550 MeV.
Densities are calculated with 
four HO frequencies 12, 16, 20, and 24 MeV for $N_\mathrm{HO}=6$, 8, 10, and 12.}
\label{fig:Li6-denr}
\end{figure}

\begin{figure}[tbp]
\centering
\includegraphics[width=\linewidth]{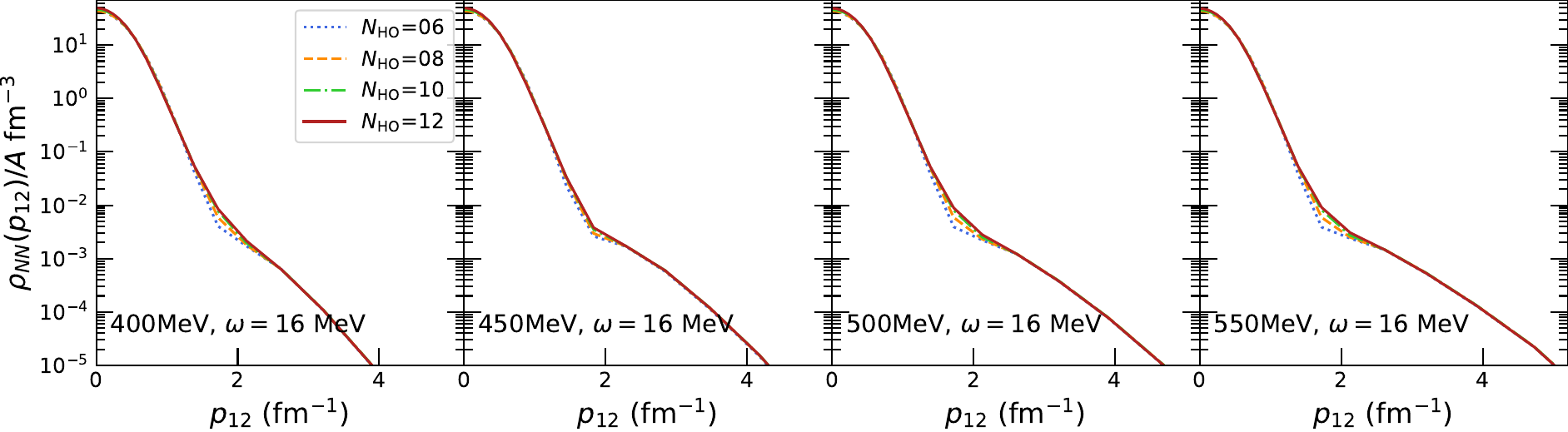}
\includegraphics[width=\linewidth]{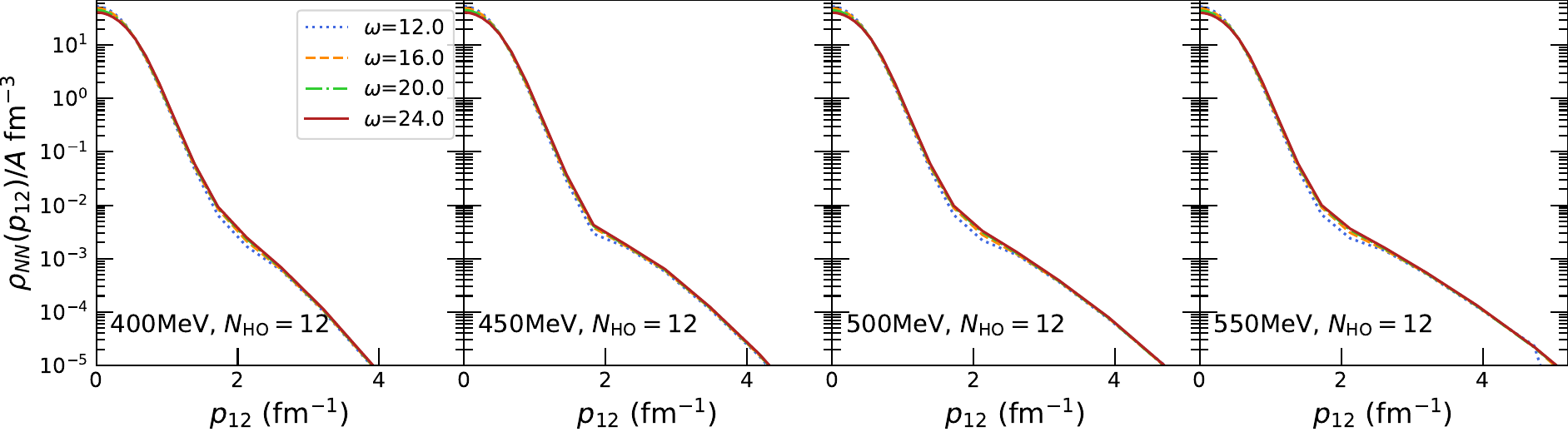}
\caption{Similar to Fig.~\ref{fig:Li6-denp}, but for $^6$He.}
\label{fig:He6-denp}
\end{figure}
    
\begin{figure}[tbp]
\centering
\includegraphics[width=\linewidth]{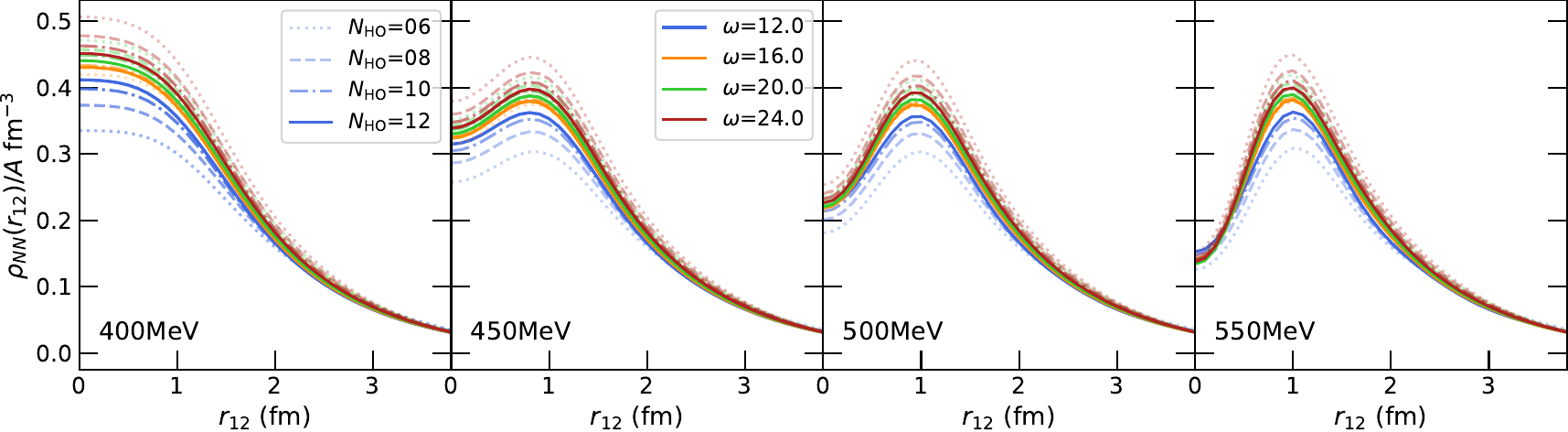}
\caption{Similar to Fig.~\ref{fig:Li6-denr}, but for $^6$He.}
\label{fig:He6-denr}
\end{figure}

\begin{figure}[tbp]
\centering
\includegraphics[width=0.9\linewidth]{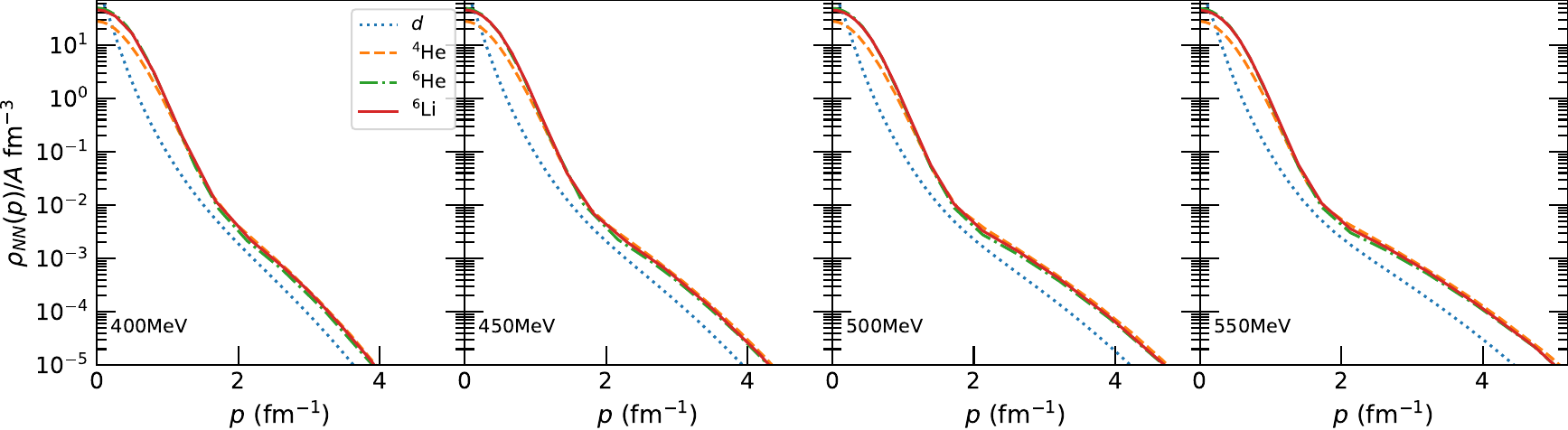}
\includegraphics[width=0.9\linewidth]{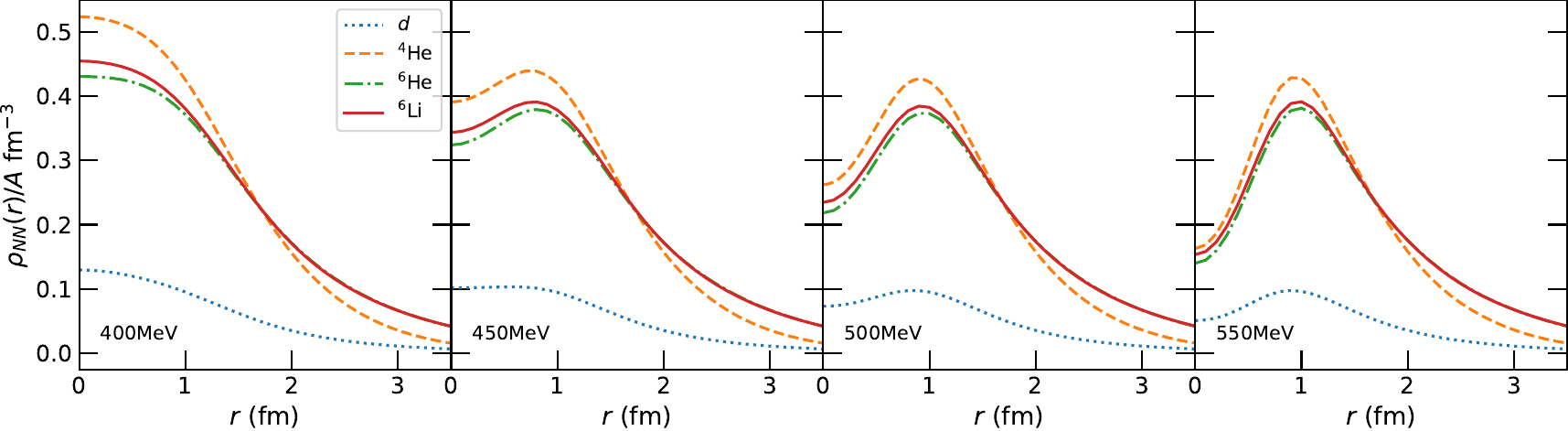}
\caption{Two-body relative densities in momentum (top panel) and coordinate (bottom panel) space of the nuclei in question. Noting in this figure the densities are scaled by the factor of $1/A$.}
\label{fig:den-all}
\end{figure}

For the $p$-shell nuclei $^{6}$Li and $^{6}$He, 
their two-body relative densities calculated using the SRG-evolved basis are shown in 
Figs. [\ref{fig:Li6-denp}, \ref{fig:Li6-denr}] and
Figs. [\ref{fig:He6-denp}, \ref{fig:He6-denr}], respectively.
In these calculations, the SRG parameter is taken to be 1.88 $\mathrm{fm}^{-1}$ for all SMS interaction with different momentum cutoffs.
For the densities in momentum space, we find that
high-momentum parts are almost independent on the HO basis size
and HO frequencies, and small differences only appear around 2 fm$^{-1}$. This is fortunate when 
studying the high-momentum 
behavior as done in this work. 
After making the Fourier transformation,
the corresponding densities in $r$-space show 
tiny dependence on the $N_\mathrm{HO}$ and $\omega$.
For the $r$-space densities, we find that the densities
for $\omega=12,20,24$ MeV tend to be converged
to the one of $\omega=16$ MeV with $N_\mathrm{HO}$ and 
those with $\omega=16$ are almost independent of the 
$N_\mathrm{HO}$. 
We have already shown that the densities' tails are almost independent of the SRG parameters in the larger momentum region.
Therefore, in Fig. \ref{fig:den-all}, we show
the densities for $d,\ ^{4,6}\mathrm{He}$, and $^6$Li in momentum and coordinate space for SMS interaction for the four available momentum cutoffs from $\Lambda_N=400$~MeV to $550$~MeV. It is clear that for all systems, the repulsive core, reflected in the tail part in $p$-space and the short-range part in $r$-space,
becomes stronger with increasing $\Lambda_N$.  The behavior in p-space  is  similar for $p_{12} > 2.5\ \mathrm{fm}^{-1}$ for the different nuclei (with the exception of the deuteron). At 
the same time, there is a significant dependence of the densities on $\Lambda_N$. 

\begin{figure}[tbp]
\centering
\includegraphics[width=0.9\linewidth]{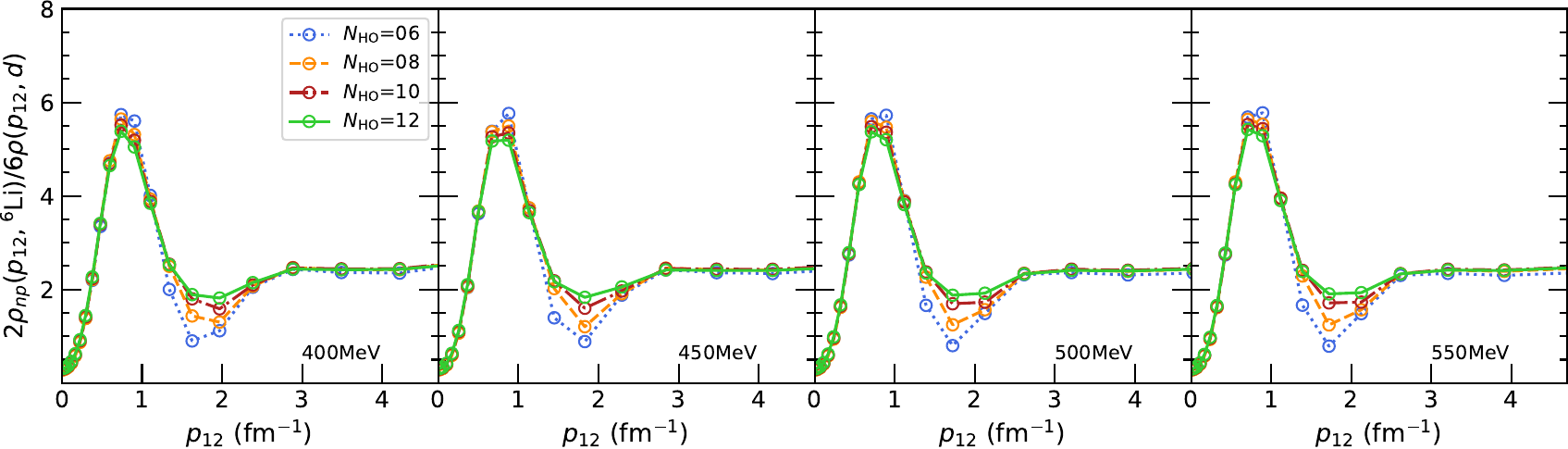}
\includegraphics[width=0.9\linewidth]{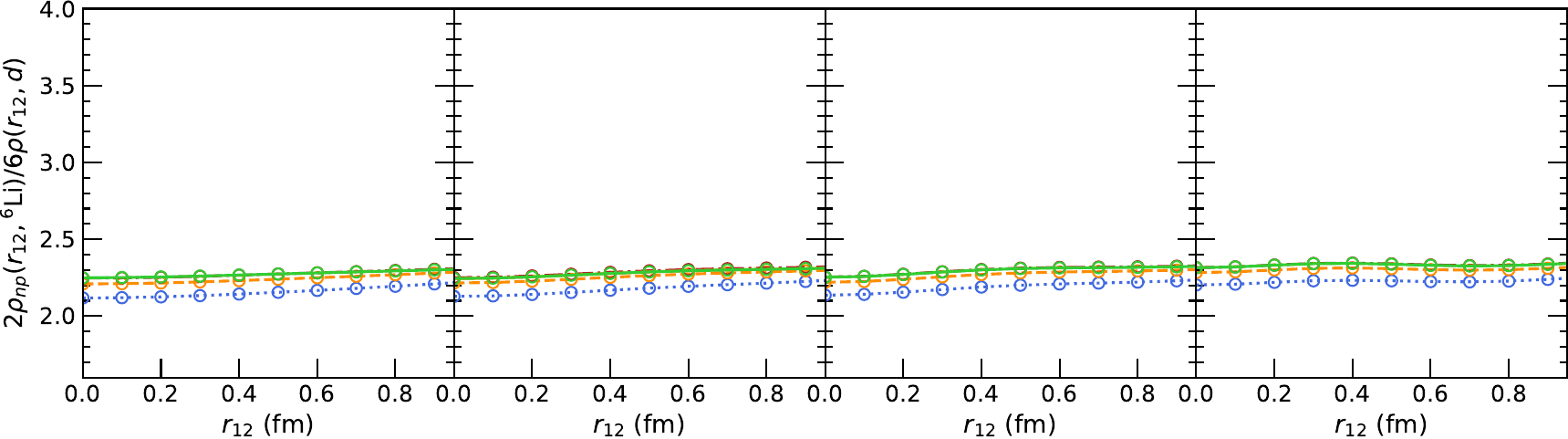}
\caption{High-momentum and short-range behavior of the ratio 
$\frac{2\rho_{NN}(^{6}\mathrm{Li})}{4\rho(d)}$ for chiral forces
SMS N$^{4}$LO$^{+}$ + N$^{2}$LO with different momentum cutoffs
($\Lambda_N=400$, 450, 500, and 550 MeV).
The two-body relative densities of $^{6}\mathrm{Li}$ from J-NCSM 
calculations with
the large HO model space $N_\mathrm{HO}=6,8,10,12$ and
four HO frequencies $\omega=16.0$ MeV.
The interactions are evolved with the flow parameter 1.88 fm$^{-1}$.
The ratios of the $^{6}$Li densities for 
the $np$ with $S=1$ channel (circles),
the $np$ channel (crosses), 
and all channels (squares)
to the deuteron density are presented.}
\label{fig:a2-Li6-d-Nho}
\end{figure}

\begin{figure}[tbp]
\centering
\includegraphics[width=0.9\linewidth]{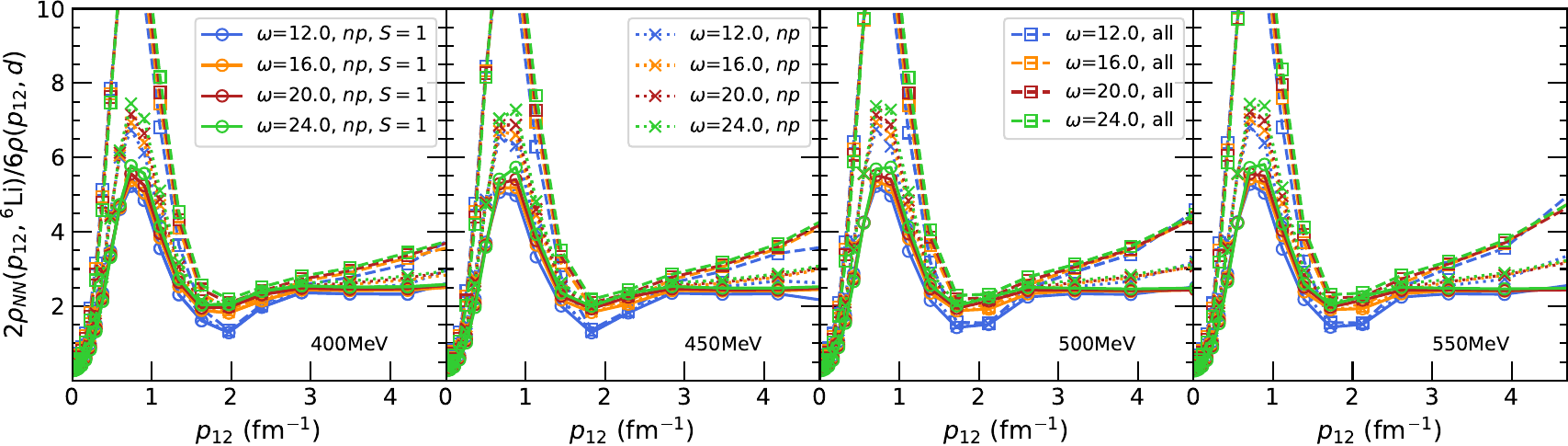}
\includegraphics[width=0.9\linewidth]{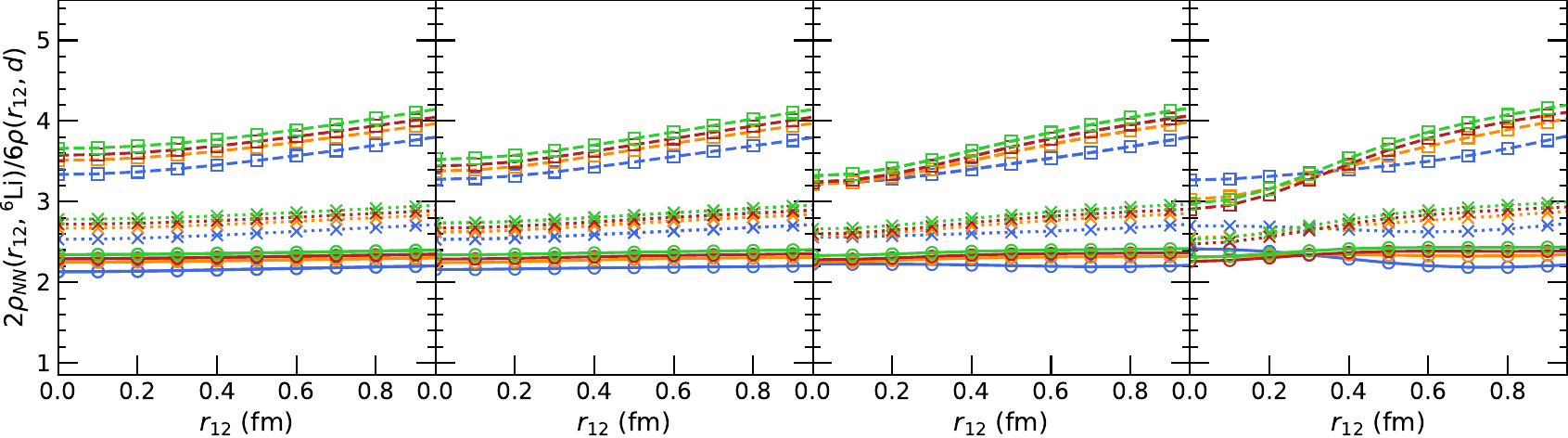}
\caption{High-momentum and short-range behavior of the ratio 
$\frac{2\rho_{NN}(^{6}\mathrm{Li})}{4\rho(d)}$ for chiral forces
SMS N$^{4}$LO$^{+}$ + N$^{2}$LO with different momentum cutoffs
($\Lambda_N=400$, 450, 500, and 550 MeV).
The two-body relative densities of $^{6}\mathrm{Li}$ from J-NCSM 
calculations with
the large HO model space $N_\mathrm{HO}=12$ and
four HO frequencies $\omega=12$, 16, 20, and 24 MeV.
The interactions are evolved with the flow parameter 1.88 fm$^{-1}$.
The ratios of the $^{6}$Li densities for 
the $np$ with $S=1$ channel (circles),
the $np$ channel (crosses), 
and all channels (squares)
to the deuteron density are presented.}
\label{fig:a2-Li6-d-p}
\end{figure}

\begin{figure}[tbp]
\centering
\includegraphics[width=0.9\linewidth]{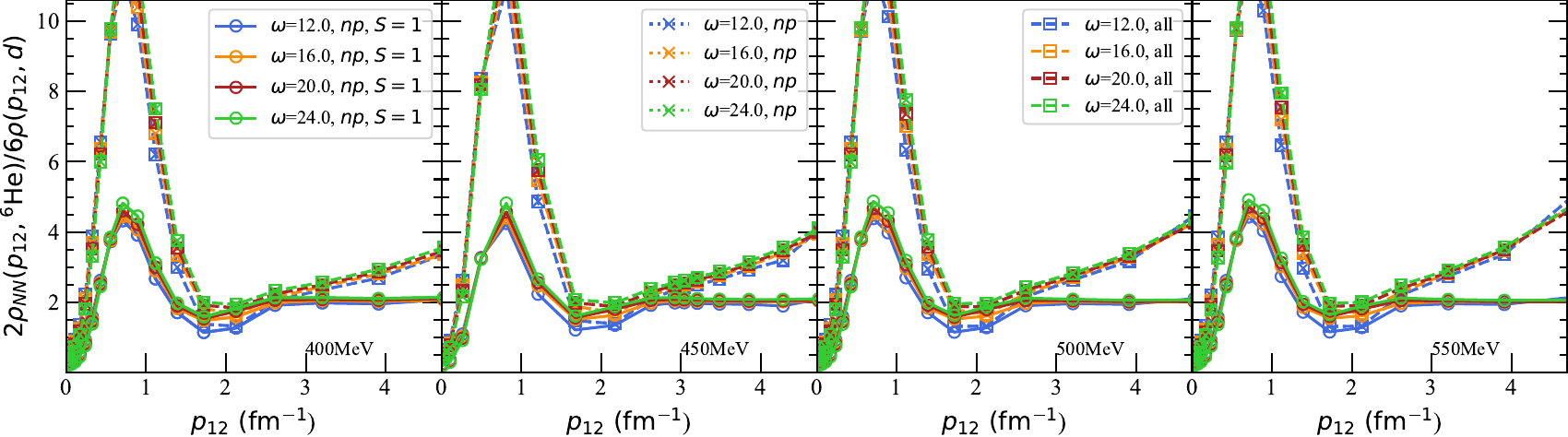}
\includegraphics[width=0.9\linewidth]{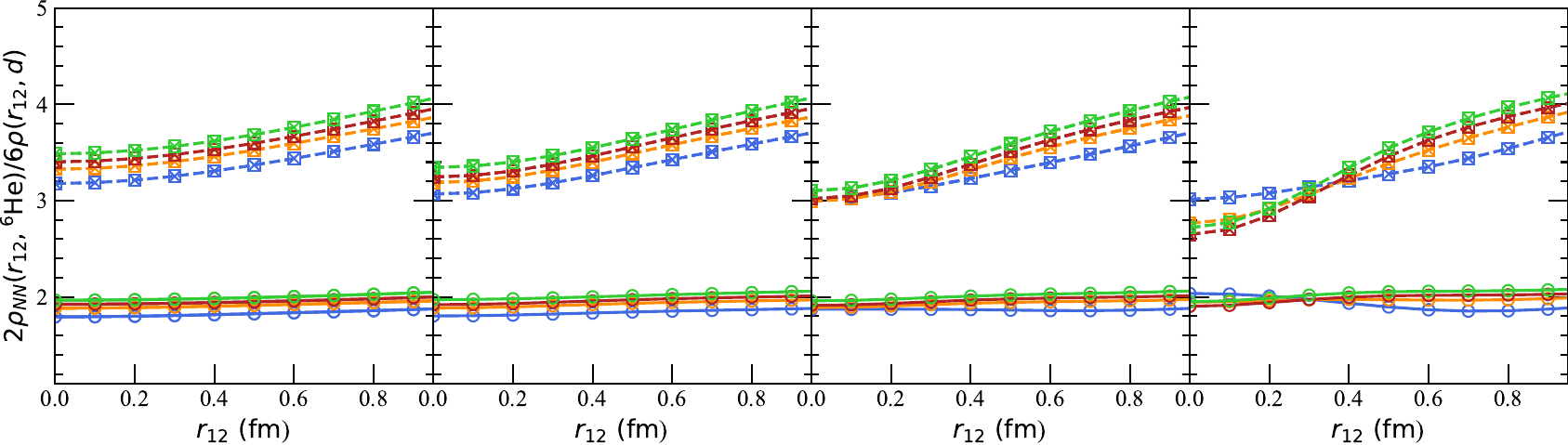}
\caption{Similar to Fig.~\ref{fig:a2-Li6-d-p}, but for $^6$He.}
\label{fig:a2-He6-d}
\end{figure}

\begin{figure}[tbp]
\centering
\includegraphics[width=0.48\linewidth]{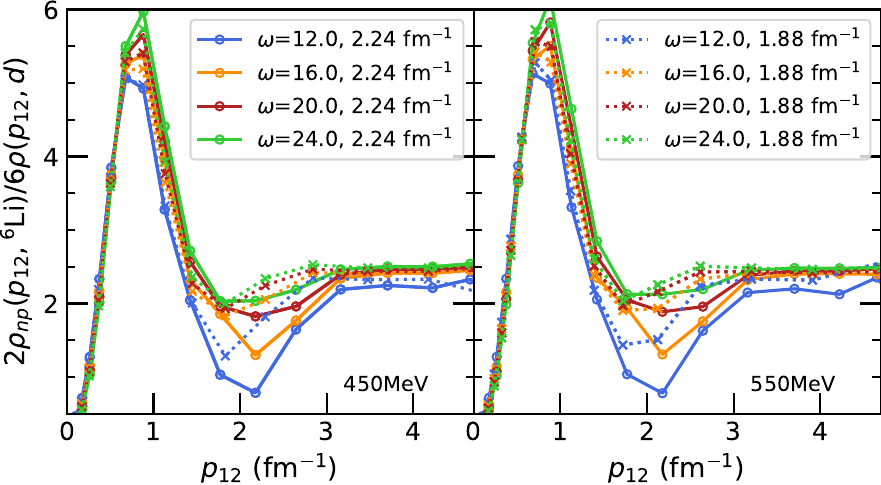}
\includegraphics[width=0.48\linewidth]{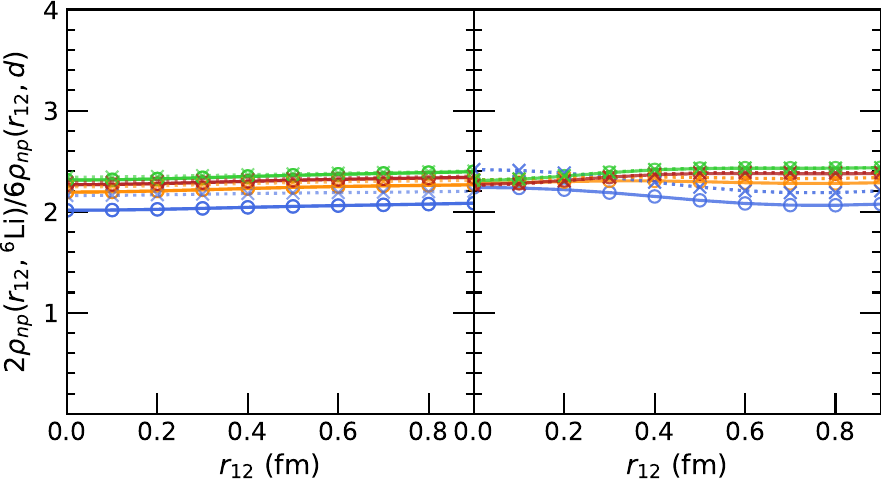}
\caption{SRG flow parameter independence of the high-momentum
and short-range behavior for $2\rho_{NN}({^{6}\mathrm{Li}})/6\rho(d)$
for the $np$-chanal with $S=1$.
The two-body relative densities of $^{6}\mathrm{Li}$ from J-NCSM 
calculations with
the large HO model space $N_\mathrm{HO}=12$ and
four HO frequencies $\omega=12$, 16, 20, and 24 MeV.
Two flow parameters 1.88 and 2.24 fm$^{-1}$ are used for $\Lambda_N=450$ and 550 MeV.}
\label{fig:a2-Li6-d-srg}
\end{figure}

\begin{table}[tbp]
\centering
    \caption{Density ratio to the deuteron for the high-momentum and short-range parts
    for SMS N$^{4}$LO$^{+}$ + N$^{2}$LO with different momentum cutoffs
    ($\Lambda_{N}=400, 450, 500$, and 550 MeV). Results 
    for flow parameters $1.88$ and $2.24$~fm$^{-1}$ are compaed 
    to the bare result where available.}
    \begin{tabular}{ccccc} 
    \toprule
         &  400 MeV&  450 MeV& 500 MeV &550 MeV \\ 
    \midrule
    \multicolumn{5}{l}{$^{4}\mathrm{He}/d$, $S=1$ $np$ channel in $p$-space} \\
    1.88 fm$^{-1}$& 2.58-2.65&  2.55-2.61&  2.52-2.55 &2.49-2.53 \\ 
    2.24 fm$^{-1}$& 2.60-2.66&  2.55-2.60&  2.52-2.59 &2.47-2.53 \\ 
    bare&           2.68-2.75&  2.64-2.86&  2.67-2.80 &2.81-2.87 \\ 
    \midrule
    \multicolumn{5}{l}{$^{4}\mathrm{He}/d$, $S=1$ $np$ channel in $r$-space} \\
    1.88 fm$^{-1}$& 2.42&  2.40&  2.39   & 2.38 \\ 
    2.24 fm$^{-1}$& 2.41&  2.37&  2.36   & 2.37 \\ 
    bare&           2.31&  2.28&  2.22   & 2.14 \\ 
    \midrule
    \multicolumn{5}{l}{$^{4}\mathrm{He}/d$, all channel in $r$-space} \\
    1.88 fm$^{-1}$&  4.088&  3.816&  3.646&3.218 \\ 
    2.24 fm$^{-1}$&  4.032&  3.766&  3.597&3.180 \\ 
    bare&            3.792&  3.498&  3.234&2.705 \\ 
    \midrule
    \multicolumn{5}{l}{$^{6}\mathrm{Li}/d$, $S=1$ $np$ channel in $p$-space} \\
    1.88 fm$^{-1}$&  2.42-2.47&  2.41-2.46&  2.39-2.44&2.41-2.46\\ 
    2.24 fm$^{-1}$& 2.39-2.53         &   2.42-2.50&  2.32-2.48         &2.37-2.41\\ 
    \midrule
    \multicolumn{5}{l}{$^{6}\mathrm{Li}/d$, $S=1$ $np$ channel in $r$-space} \\
    1.88 fm$^{-1}$&  2.25  &  2.24&  2.25 &2.31\\ 
    2.24 fm$^{-1}$&  2.18  &  2.21&  2.20 &2.28\\ 
    \midrule
    \multicolumn{5}{l}{$^{6}\mathrm{He}/d$, $S=1$ $np$ channel in $p$-space} \\
    1.88 fm$^{-1}$&  2.03-2.07&  2.02-2.04&  1.95-2.10&2.00-2.04 \\ 
    2.24 fm$^{-1}$& 2.04-2.08 &  2.02-2.06&           &2.00-2.05   \\ 
     \midrule
    \multicolumn{5}{l}{$^{6}\mathrm{He}/d$, $S=1$ $np$ channel in $r$-space} \\
    1.88 fm$^{-1}$& 1.88&  1.89 &  1.89& 1.94\\ 
    2.24 fm$^{-1}$& 1.83    &  1.85 &    1.85  &1.87 \\     
    \bottomrule
    \end{tabular}
    \label{tab:tab1}
\end{table}

\section{Scaling factor \texorpdfstring{$a_2(A,d)$}{a2(A,d)} from J-NCSM calculations}
\label{sec:s_iv}
The scaling factors are extracted using the ratios of densities in the high-momentum part and the zero-range part. Before we go to the exact values of $a_2(A,d)$, we need to examine the residual dependence on the basis parameters: $N_\mathrm{HO}$ and $\omega$.
Figure~\ref{fig:a2-Li6-d-Nho} shows 
the ratio $\frac{2\rho_{NN}(^{6}\mathrm{Li})}{4\rho(d)}$ for four SMS interactions with
$\omega=16.0$ MeV for different basis sizes. 
It is clear that, for the high momentum region, the ratio
is almost unchanged with $N_\mathrm{HO}$. Also for the zero-range part, the ratio is well converged with increasing the basis size. Therefore, in the following discussions, we focus on the results with $N_\mathrm{HO}=12$. 
In Fig. \ref{fig:a2-Li6-d-p}, we show 
the same ratio $\frac{2\rho_{NN}(^{6}\mathrm{Li})}{4\rho(d)}$ dependent on $\omega$ keeping 
$N_\mathrm{HO}=12$. Although the range of $\omega=12,\ldots, 24$~MeV 
is rather large, one sees that the high-momentum part is almost 
independent of $\omega$. Using r-space, the ratio is slightly more 
dependent on $\omega$ at short distances, still, the $\omega=16.0$ and $20.0$~MeV results are almost identical. This covers the range 
of optimal values for $\omega$. Based on these observations, we 
determine the ratio of the large momentum parts using the densities with $N_\mathrm{HO}=12$ and $\omega=16.0$ (20.0) MeV. 
Same conclusions can also be obtained for $^{6}\mathrm{He}$ as can be seen in Fig. \ref{fig:a2-He6-d}.

For the ratio to the deuteron, we generally observe that the 
independence of the momentum or the distance realized to a very high accuracy only for the $S=1$ np channel. We will see below that 
building the ratio to $^4$He is important to extend to other pairs. 
Therefore, we finally confirm that our conclusions are independent of the adopted SRG flow parameter only for these most important $S=1$ pairs. This is shown in Fig. \ref{fig:a2-Li6-d-srg} for two SMS interaction. The SRG parameter dependence is negligible for the 
optimal $\omega$. Small deviations for $\omega=12$~MeV can be 
explained due to convergence with respect to $N_{\rm HO}$, which 
is less good for this frequency. In summary, the analysis shows 
that the ratios can be extracted independently of the 
SRG parameter and model space from the NCSM wave functions after the SRG back transformations have been applied.

The resulting scaling factors are summarized in Tab.~\ref{tab:tab1}. 
For the final extraction  in momentum space, we use the maximum and minimum of the ratios in the momentum interval $(3.3,\ 5.0)$ fm$^{-1}$ which also gives a  measure of the 
remaining uncertainty. We refrain from using the values 
$p_{12} > 5\ \mathrm{fm}^{-1}$ because the densities in these regions are small and they sometimes show unphysical oscillations so that 
the ratios become numerically unreliable. We also only extract the 
ratio for $S=1$ np channels in momentum space because the residual 
momentum dependence for the other channels is significant. 
However, we include ratios to other channels from r-space 
for completeness. 

In $r$-space, the densities are obtained by performing the Fourier transformation on the momentum space wave functions, i.e. the 
transformed basis functions of Eq.~\eqref{eq:hoevolvedwf}. We can then get the values at $r=0$ and calculate the ratios. We use the values with $\omega=16.0$ and $20.0$~MeV to determine the ratios. 

Although we could show that these ratios are independent of model spaces, frequency and SRG parameter, there is a small but systematic 
deviation of the r-space and p-space ratio. We will see below 
that this difference is much smaller for the ratio to $^4$He. 
We therefore believe that such a systematic difference could be 
related to total momentum of the considered pairs. Clearly, 
for the deuteron, only pairs at rest are considered whereas 
the densities for other nuclei involve averaging over the 
total pair momentum. We find that for ratios to the deuteron 
the extraction in r-space and p-space differ by approximately 
10\%. This should be kept in mind, when comparing to other 
approaches.

\section{Scaling factor \texorpdfstring{$a_2(A,{}^{4}\mathrm{He})$}{a2(A,4-He)} from J-NCSM calculations}

\begin{figure}[tbp]
\centering
\includegraphics[width=\linewidth]{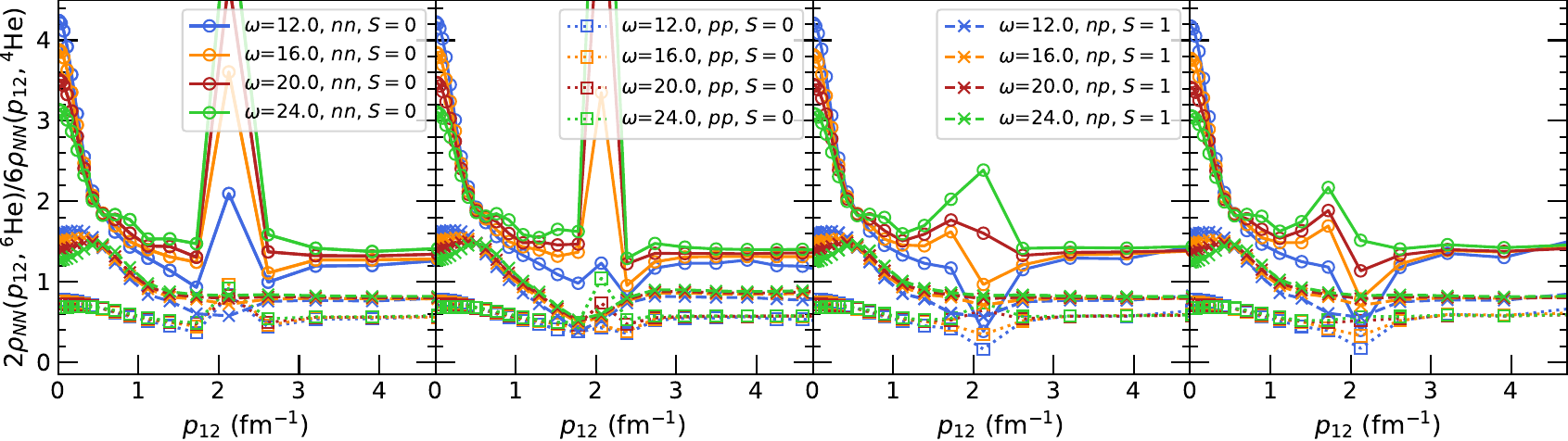}
\includegraphics[width=\linewidth]{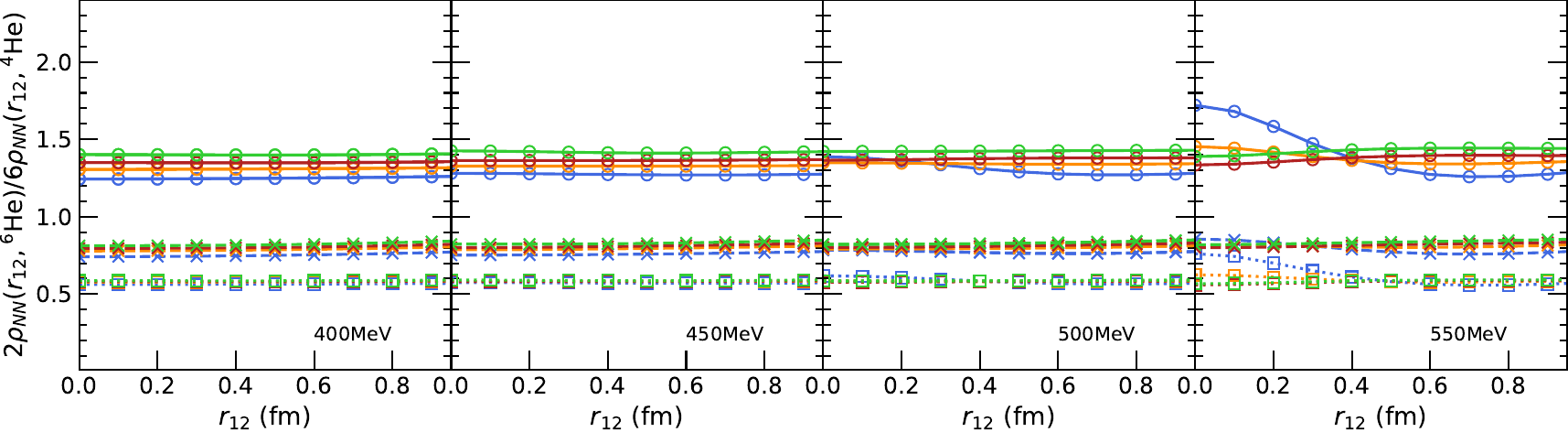}
\caption{High-momentum and short-range behavior of the ratio 
$\frac{4\rho_{NN}(^{6}\mathrm{He})}{6\rho_{NN}(^{4}\mathrm{He})}$ for chiral forces
SMS N$^{4}$LO$^{+}$ + N$^{2}$LO with different momentum cutoffs
($\Lambda_N=400$, 450, 500, and 550 MeV).
The two-body relative densities of $^{6}\mathrm{He}$ from J-NCSM 
calculations with
the large HO model space $N_\mathrm{HO}=12$ and
four HO frequencies $\omega=12$, 16, 20, and 24 MeV is employed.
The interactions are evolved with the flow parameter 1.88 fm$^{-1}$.
The ratio for both the $nn$-channel with $S=0$ (labeled as circles),
$pp$-channel with $S=0$ (labeled as squares), and
$np$-channel with $S=1$ (labeled as crosses) are presented.}
\label{fig:a2-He6-He4}
\end{figure}

\begin{figure}[tbp]
\centering
\includegraphics[width=\linewidth]{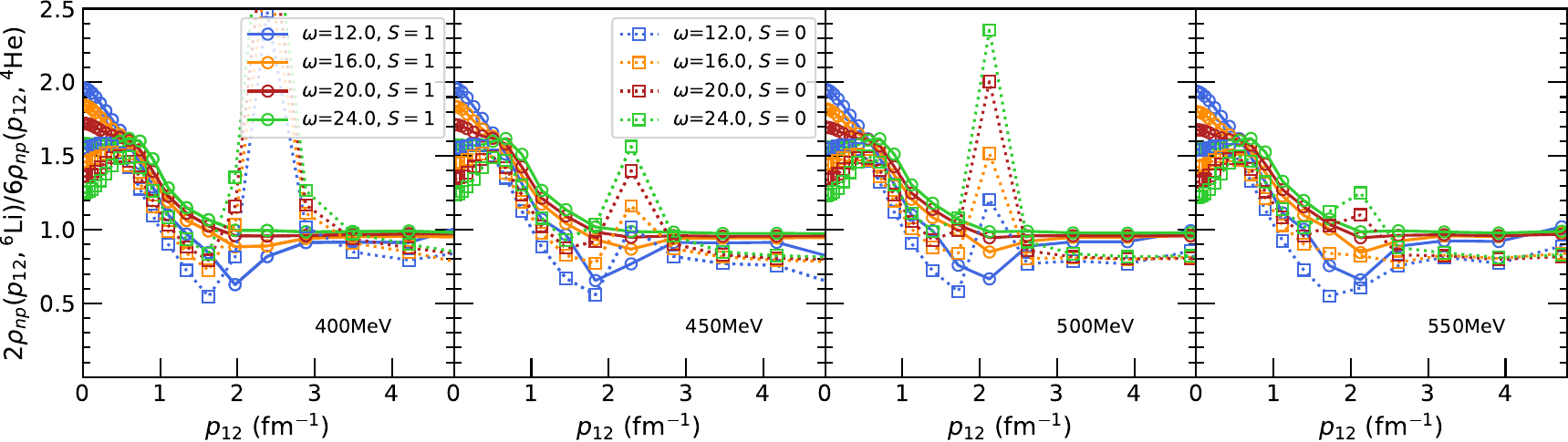}
\includegraphics[width=\linewidth]{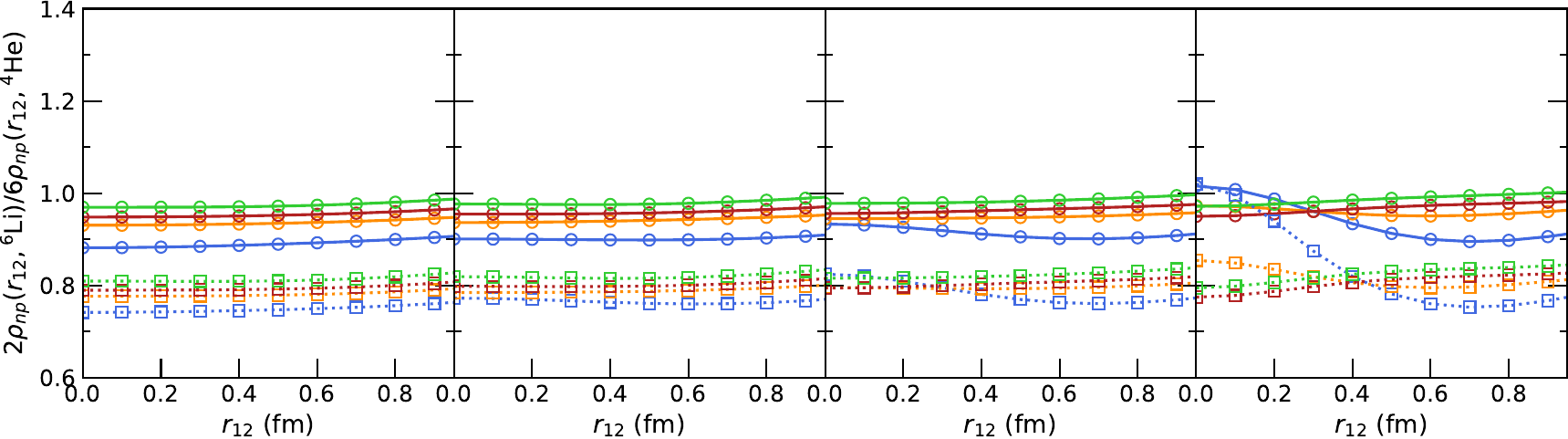}
\caption{Similar to Fig.~\ref{fig:a2-He6-He4} for the ratio of $^6$Li to $^4$He. In order to avoid overlapping results, only  the ratio for  the $np$-channel with $S = 1$ (labeled as circles), 
and the $np$-channel with $S = 0$ (labeled as squares) are presented.}
\label{fig:a2-Li6-He4-np}
\end{figure}

\begin{figure}[tbp]
    \centering
    \includegraphics[width=\linewidth]{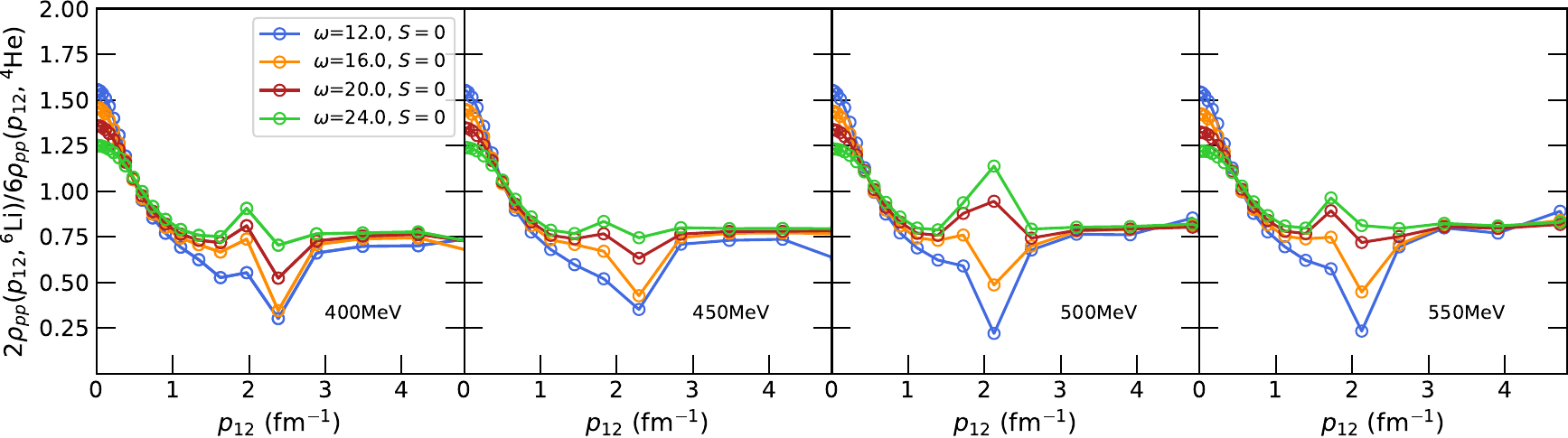}
    \includegraphics[width=\linewidth]{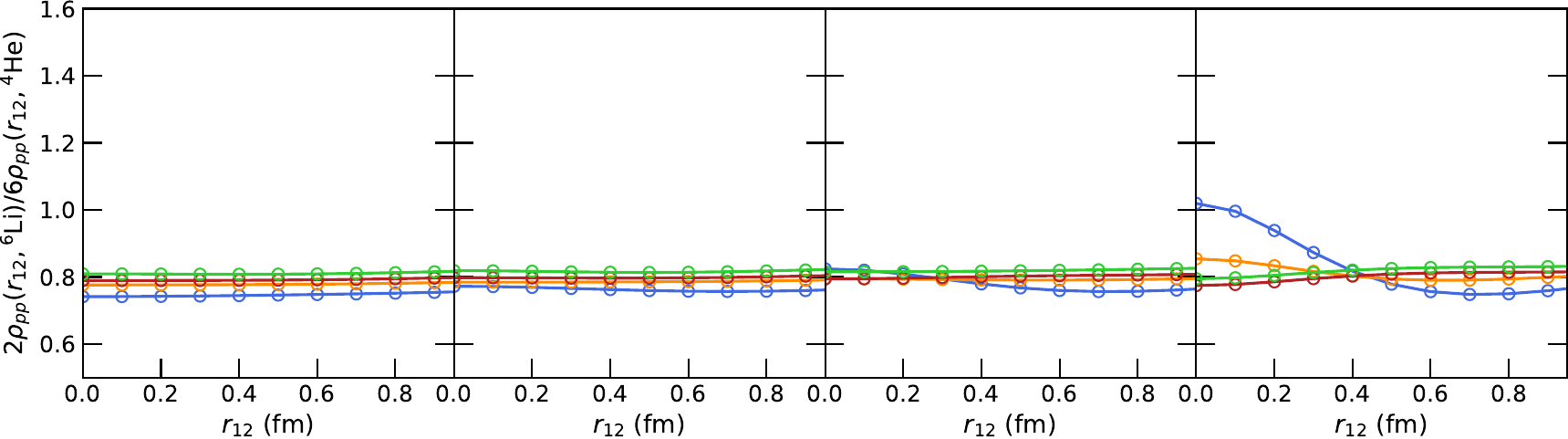}
    \caption{Similar to Fig.~\ref{fig:a2-Li6-He4-np}, but for the $pp$-channel with $S=0$.}
    \label{fig:a2-Li6-He4-pp}
\end{figure}

\begin{figure}[tbp]
    \centering
    \includegraphics[width=\linewidth]{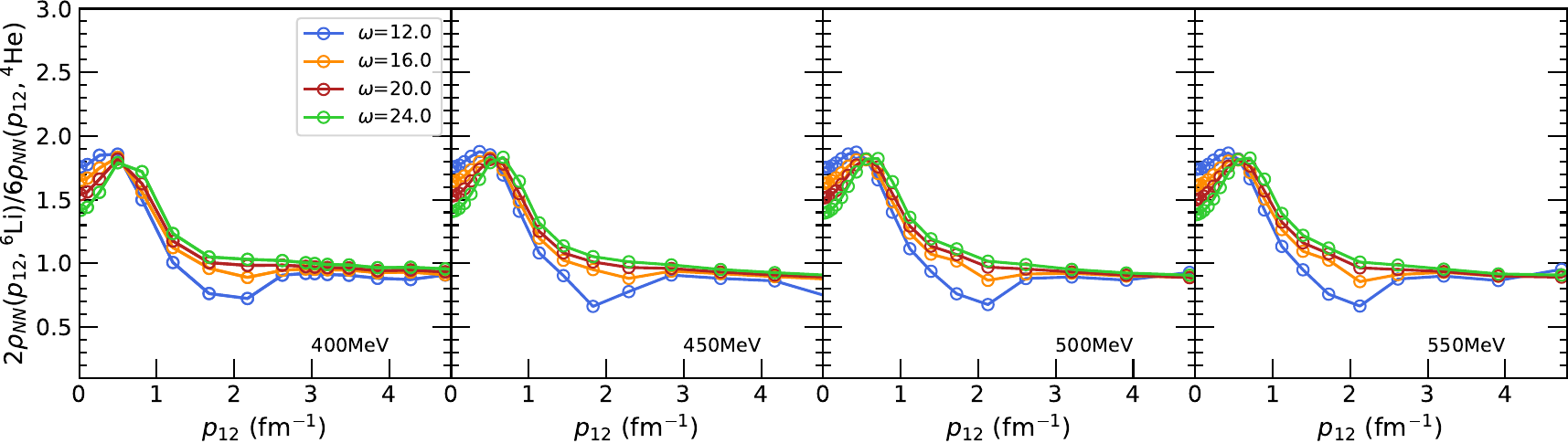}
    \includegraphics[width=\linewidth]{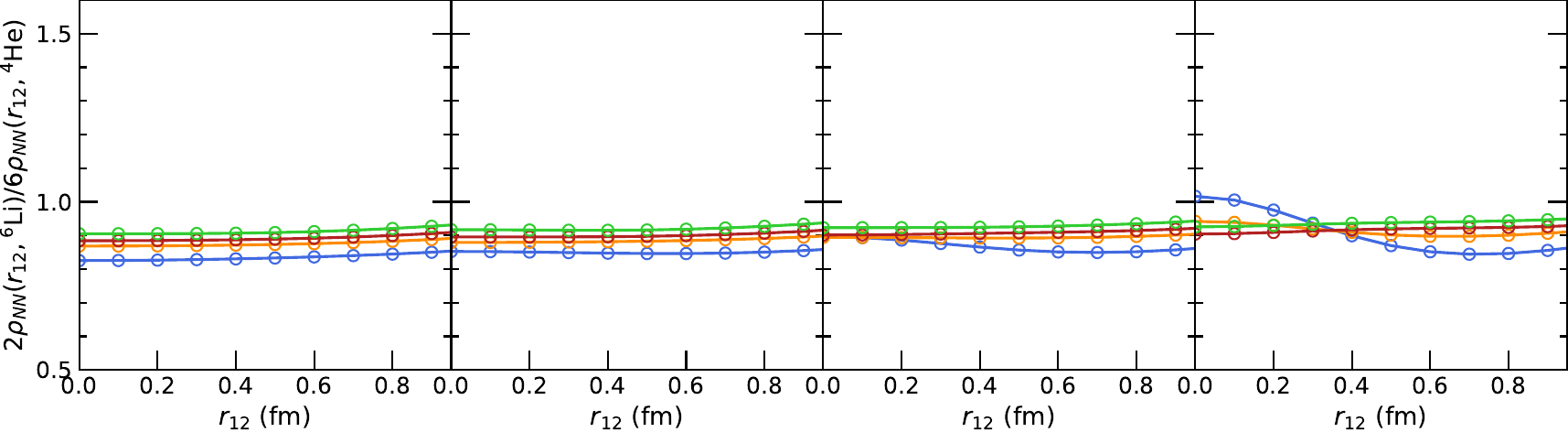}
    \caption{Similar to Fig.~\ref{fig:a2-Li6-He4-pp}, but for all
    channels.}
    \label{fig:a2-Li6-He4-all}
\end{figure}

\begin{figure}[tbp]
    \centering
    \includegraphics[width=\linewidth]{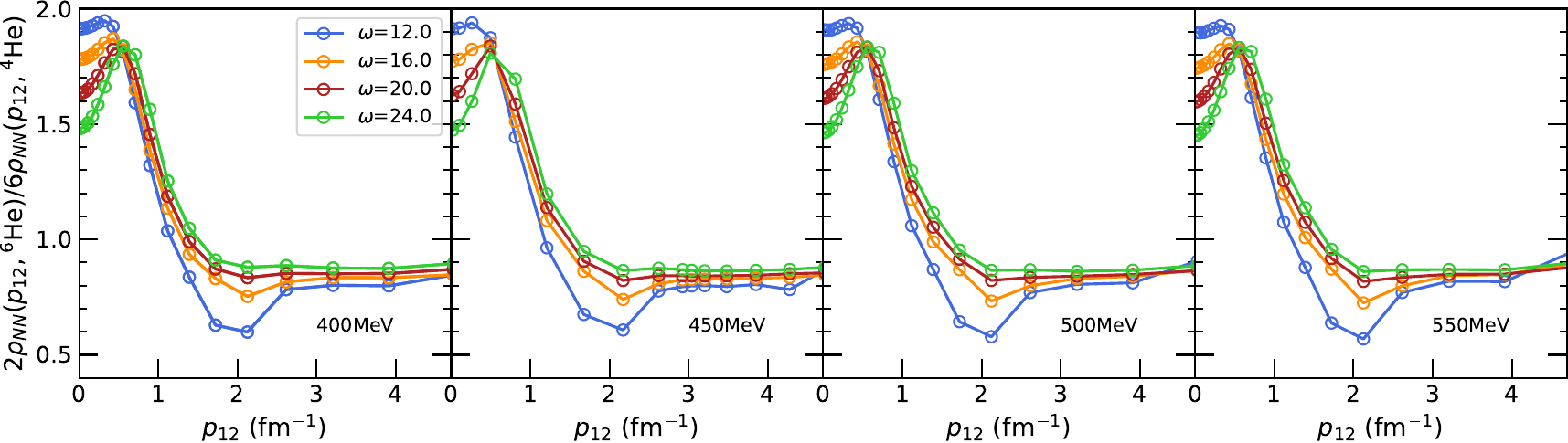}
    \includegraphics[width=\linewidth]{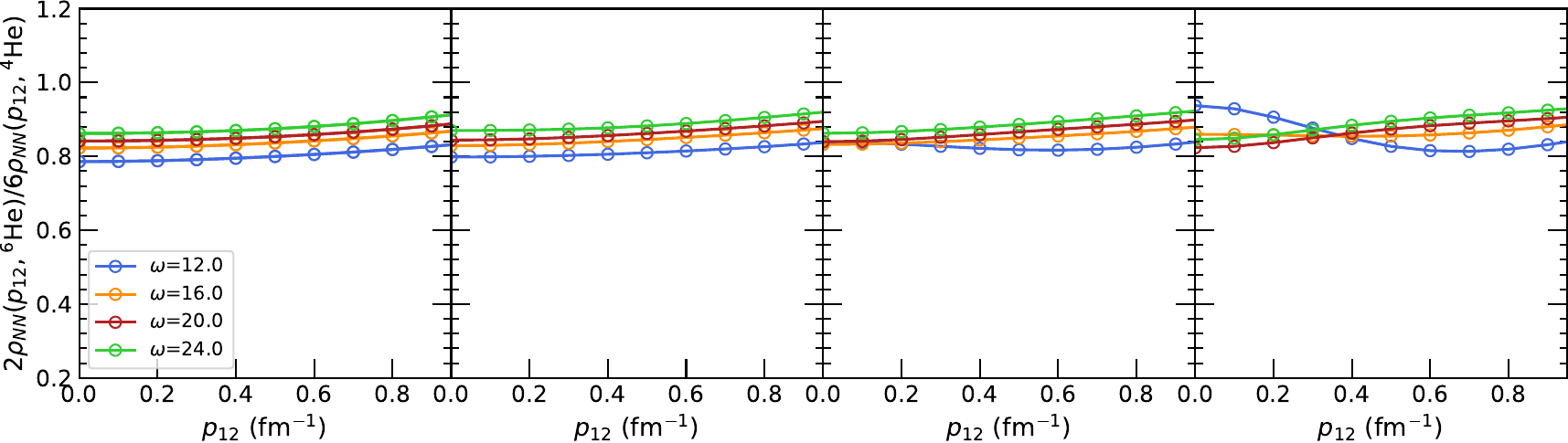}
    \caption{Similar to Fig.~\ref{fig:a2-Li6-He4-all}, but for 
    the ratio of $^6$He to $^4$He.}
    \label{fig:a2-He6-He4-all}
\end{figure}

\begin{figure}[tbp]
\centering
\includegraphics[width=0.48\linewidth]{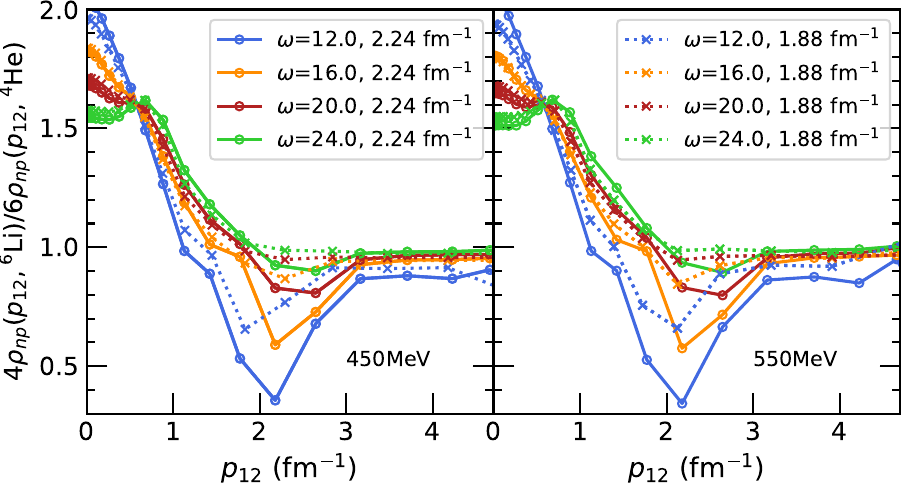}
\includegraphics[width=0.48\linewidth]{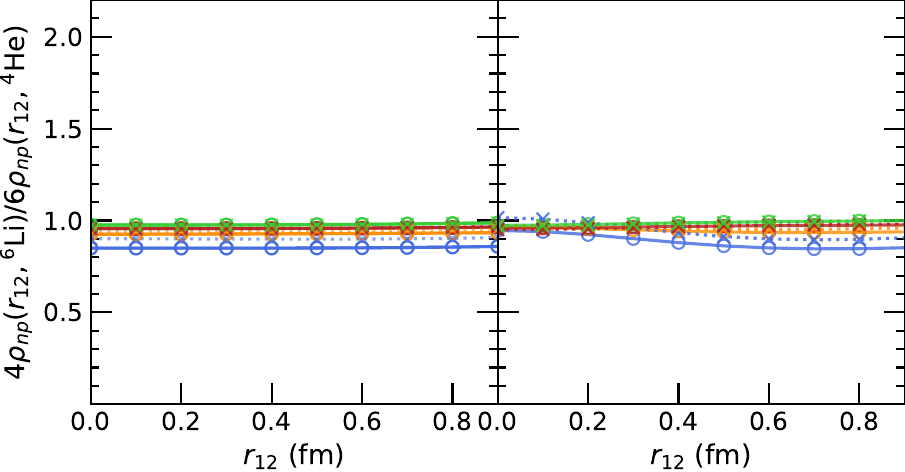}
\caption{Similar to Fig.~\ref{fig:a2-Li6-d-srg}, but for 
$4\rho_{NN}({^{6}\mathrm{Li}})/6\rho(^{4}\mathrm{He})$
in the $np$-channel with $S=1$.}
\label{fig:a2-Li6-He4-srg}
\end{figure}

\begin{figure}[tbp]
\centering
\includegraphics[width=0.48\linewidth]{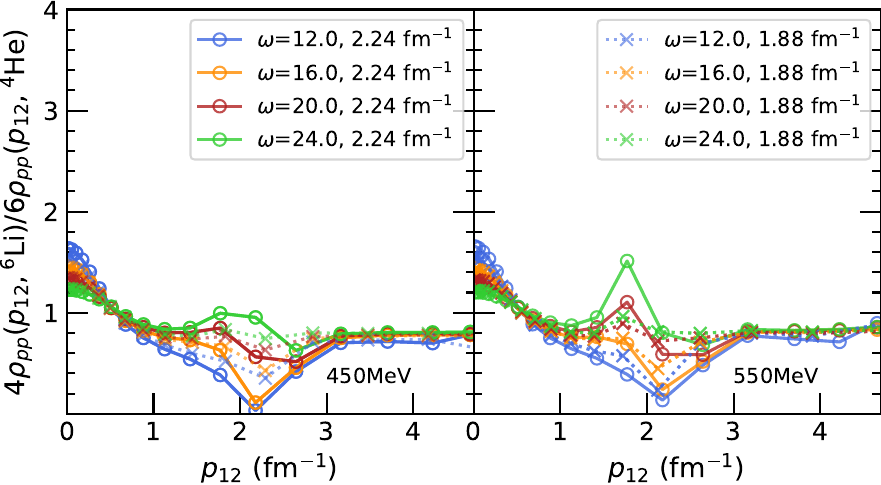}
\includegraphics[width=0.48\linewidth]{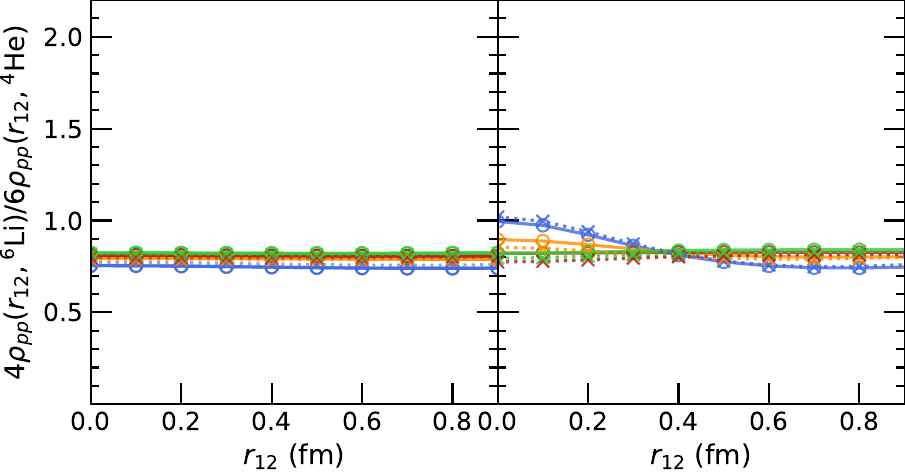}
\caption{Similar to Fig.~\ref{fig:a2-Li6-He4-srg}, 
but for $pp$-channel with $S=0$.}
\label{fig:a2-Li6-He4-srg-pp}
\end{figure}

We performed the same procedure
discussed in Sec. \ref{sec:s_iv} to extract the ratio with respect to the
$\alpha$ particle for different spin-isospin channels and found similar 
conclusions. The results for $\frac{4\rho_{NN}(^{6}\mathrm{He})}
{6\rho_{NN}(^{4}\mathrm{He})}$ and $\frac{4\rho_{NN}(^{6}\mathrm{Li})}
{6\rho_{NN}(^{4}\mathrm{He})}$ are shown in Figures~\ref{fig:a2-He6-He4}
to \ref{fig:a2-Li6-He4-srg}. The first Figures~\ref{fig:a2-He6-He4}
to \ref{fig:a2-Li6-He4-all} confirm for 
$^6$He and $^6$Li for different two-nucleon channels that 
there is only a mild dependence on the employed basis and 
the used chiral interaction. The last two Figures~\ref{fig:a2-Li6-He4-srg} and \ref{fig:a2-Li6-He4-srg-pp} show additionally the dependence on 
the SRG parameter which is again insignificant. 
 
The obtained scaling factors are summarized in Tab.~\ref{tab:tab2} for 
the $np$ $S=1$ channel,  Tab.~\ref{tab:tab3} for the $pp$ $S=0$ channel, 
and Tab.~\ref{tab:tab4} for the total two-body density. As mentioned 
above, the ratios extracted in p- and r-space show a much better 
agreement for the ration to $^4$He which can probably be traced back to to the CM momentum \cite{CLAS:2018qpc}.

\begin{table}[tbp]
\centering
    \caption{Density ratio to $^{4}\mathrm{He}$ for the $S=1$ $np$ channel at high-momentum and short-range parts
    for bare SMS N$^{4}$LO$^{+}$ + N$^{2}$LO interaction with different momentum cutoffs
    ($\Lambda_{N}=400, 450, 500$, and 550 MeV).}
    \begin{tabular}{ccccc} 
    \toprule
         &  400 MeV&  450 MeV& 500 MeV &550 MeV \\ 
    \midrule
    \multicolumn{5}{l}{$^{6}\mathrm{Li}/^{4}\mathrm{He}$,  $p$-space} \\
    1.88 fm$^{-1}$&  0.94-0.96&  0.94-0.95&  0.92-0.93& 0.95-0.98\\ 
    2.24 fm$^{-1}$&  0.93-0.96&  0.94-0.96&  0.95-0.97& 0.95-0.96\\ 
    \midrule
    \multicolumn{5}{l}{$^{6}\mathrm{Li}/^{4}\mathrm{He}$,  $r$-space} \\
    1.88 fm$^{-1}$&  0.93  &  0.94&  0.94  &0.97\\ 
    2.24 fm$^{-1}$&  0.89  &0.92&    0.94  &0.96\\ 
    \midrule
    \multicolumn{5}{l}{$^{6}\mathrm{He}/^{4}\mathrm{He}$,  $p$-space} \\
    1.88 fm$^{-1}$& 0.79-0.80&  0.78-0.79&  0.79-0.80 &0.79-0.81 \\ 
    2.24 fm$^{-1}$& 0.79-0.82& 0.79-0.80&  0.79-0.82  &0.78-0.81   \\ 
     \midrule
    \multicolumn{5}{l}{$^{6}\mathrm{He}/^{4}\mathrm{He}$,  $r$-space} \\
    1.88 fm$^{-1}$& 0.78 &  0.79 &  0.79& 0.82\\ 
    2.24 fm$^{-1}$& 0.76 &  0.77 &  0.78&0.79 \\     
    \bottomrule
    \end{tabular}
    \label{tab:tab2}
\end{table}

\begin{table}[tbp]
\centering
    \caption{Density ratio to $^{4}\mathrm{He}$ for the $S=0$ $pp$ channel at high-momentum and short-range parts
    for bare SMS N$^{4}$LO$^{+}$ + N$^{2}$LO interaction with different momentum cutoffs
    ($\Lambda_{N}=400, 450, 500$, and 550 MeV).}
    \begin{tabular}{ccccc} 
    \toprule
         &  400 MeV&  450 MeV& 500 MeV &550 MeV \\ 
    \midrule
    \multicolumn{5}{l}{$^{6}\mathrm{Li}/^{4}\mathrm{He}$,  $p$-space} \\
    1.88 fm$^{-1}$&  0.74-0.75&  0.76-0.77&  0.78-0.81& 0.80-0.83\\ 
    2.24 fm$^{-1}$&  0.71-0.75&   0.76-0.77& 0.76-0.81&0.77-0.81\\ 
    \midrule
    \multicolumn{5}{l}{$^{6}\mathrm{Li}/^{4}\mathrm{He}$,  $r$-space} \\
    1.88 fm$^{-1}$&  0.78  &  0.77&  0.79  &0.85\\ 
    2.24 fm$^{-1}$&  0.77   &0.78&  0.79 &0.86\\ 
    \midrule
    \multicolumn{5}{l}{$^{6}\mathrm{He}/^{4}\mathrm{He}$,  $p$-space} \\
    1.88 fm$^{-1}$& 0.55-0.56& 0.57-0.58&  0.58-0.60 &0.59-0.61 \\ 
    2.24 fm$^{-1}$&0.55-0.61 &0.54-0.57&  0.57-0.62          &0.57-0.60  \\ 
     \midrule
    \multicolumn{5}{l}{$^{6}\mathrm{He}/^{4}\mathrm{He}$,  $r$-space} \\
    1.88 fm$^{-1}$& 0.58 &  0.58 &  0.58& 0.62\\ 
    2.24 fm$^{-1}$& 0.59     &  0.57 & 0.60  &0.58\\     
    \bottomrule
    \end{tabular}
    \label{tab:tab3}
\end{table}

\begin{table}[tbp]
\centering
    \caption{Density ratio to $^{4}\mathrm{He}$ for all channels at high-momentum and short-range parts
    for bare SMS N$^{4}$LO$^{+}$ + N$^{2}$LO interaction with different momentum cutoffs
    ($\Lambda_{N}=400, 450, 500$, and 550 MeV).}
    \begin{tabular}{ccccc} 
    \toprule
         &  400 MeV&  450 MeV& 500 MeV &550 MeV \\ 
    \midrule
    \multicolumn{5}{l}{$^{6}\mathrm{Li}/^{4}\mathrm{He}$,  $p$-space} \\
    1.88 fm$^{-1}$&  0.91-0.93&  0.88-0.90&  0.89-0.90 & 0.90-0.91\\ 
    2.24 fm$^{-1}$& 0.92-0.96 &   0.89-0.92& 0.87-0.92 & 0.88-0.91\\ 
    \midrule
    \multicolumn{5}{l}{$^{6}\mathrm{Li}/^{4}\mathrm{He}$,  $r$-space} \\
    1.88 fm$^{-1}$&  0.87  &  0.88&  0.89  &0.94\\ 
    2.24 fm$^{-1}$&  0.84  &  0.87&  0.90  &0.94\\ 
    \midrule
    \multicolumn{5}{l}{$^{6}\mathrm{He}/^{4}\mathrm{He}$,  $p$-space} \\
    1.88 fm$^{-1}$& 0.83-0.85& 0.83-0.85&  0.84-0.87 &0.85-0.88 \\ 
    2.24 fm$^{-1}$& 0.84-0.89&0.84-0.86&   0.83-0.88 &0.84-0.89  \\ 
     \midrule
    \multicolumn{5}{l}{$^{6}\mathrm{He}/^{4}\mathrm{He}$,  $r$-space} \\
    1.88 fm$^{-1}$& 0.82 &  0.83 &  0.83& 0.86\\ 
    2.24 fm$^{-1}$& 0.81 &  0.82 &  0.83&0.82\\     
    \bottomrule
    \end{tabular}
    \label{tab:tab4}
\end{table}



\putbib
\end{bibunit}

\end{document}